\documentclass[%
reprint,
superscriptaddress,
nofootinbib,
 amsmath,amssymb,
aps,
prx,
floatfix,
]{revtex4-2}

\usepackage{dcolumn}
\usepackage{siunitx}
\usepackage{graphicx}
\usepackage{times}
\usepackage{braket}
\usepackage{multirow}
\usepackage{rotating}
\usepackage{tabularray}
\usepackage{hyperref}
\newcolumntype{x}[1]{>{\raggedright\arraybackslash}p{#1}}
\newcolumntype{M}[1]{>{\centering\arraybackslash}m{#1}}

\begin{document}

\title{Metropolitan-scale heralded entanglement of solid-state qubits}

\author{A.J. Stolk}\altaffiliation{These authors contributed equally to this work}
\author{K.L. van der Enden}\altaffiliation{These authors contributed equally to this work}
\author{M.-C. Slater}
\author{I. te Raa-Derckx}
\author{P. Botma}
\author{J. van Rantwijk}
\affiliation{%
\vspace{0em}QuTech \& Kavli Institute of Nanoscience, Delft University of Technology, 2628 CJ Delft, The Netherlands
}

\author{J.J.B. Biemond}
\author{R.A.J. Hagen}
\author{R.W. Herfst}
\author{W.D. Koek}
\author{A.J.H. Meskers}
\author{R. Vollmer}
\author{E.J. van Zwet}
\affiliation{%
\vspace{0em}QuTech \& Kavli Institute of Nanoscience, Delft University of Technology, 2628 CJ Delft, The Netherlands
}
\affiliation{%
\vspace{0em}Netherlands Organisation for Applied Scientific Research (TNO), P.O. Box 155, 2600 AD Delft, The Netherlands
}
\author{M. Markham}
\author{A.M. Edmonds}
\affiliation{%
\vspace{0em}Element Six Innovation,Fermi Avenue, Harwell Oxford, Didcot, Oxfordshire OX11 0QR, United Kingdom
}

\author{J.F. Geus}

\author{F. Elsen}
\affiliation{\vspace{0em}
Fraunhofer Institute for Laser Technology ILT, 52074 Aachen, Germany
}
\affiliation{\vspace{0em}
Chair for Laser Technology, RWTH-Aachen University, 52074 Aachen, Germany
}
\author{B. Jungbluth}
\affiliation{\vspace{0em}
Fraunhofer Institute for Laser Technology ILT, 52074 Aachen, Germany
}
\author{C. Haefner}
\affiliation{\vspace{0em}
Fraunhofer Institute for Laser Technology ILT, 52074 Aachen, Germany
}
\affiliation{\vspace{0em}
Chair for Laser Technology, RWTH-Aachen University, 52074 Aachen, Germany
}

\author{C. Tresp}
\author{J. Stuhler}
\author{S. Ritter}
\affiliation{%
\vspace{0em}TOPTICA Photonics AG, Lochhamer Schlag 19, 82166 Graefelfing, Germany
}
\author{R. Hanson}
\email{Correspondence to: R.Hanson@tudelft.nl}
\affiliation{%
\vspace{0em}QuTech \& Kavli Institute of Nanoscience, Delft University of Technology, 2628 CJ Delft, The Netherlands
}

\date{\today}
\begin{abstract}
    A key challenge towards future quantum internet technology is connecting quantum processors at metropolitan scale. Here, we report on heralded entanglement between two independently operated quantum network nodes separated by 10km. The two nodes hosting diamond spin qubits are linked with a midpoint station via 25km of deployed optical fiber. We minimize the effects of fiber photon loss by quantum frequency conversion of the qubit-native photons to the telecom L-band and by embedding the link in an extensible phase-stabilized architecture enabling the use of the loss-resilient single-photon entangling protocol. By capitalizing on the full heralding capabilities of the network link in combination with real-time feedback logic on the long-lived qubits, we demonstrate the delivery of a predefined entangled state on the nodes irrespective of the heralding detection pattern. Addressing key scaling challenges and being compatible with different qubit systems, our architecture establishes a generic platform for exploring metropolitan-scale quantum networks.
\end{abstract}

\footnotetext{M.-C. S. is currently at: AIT Austrian Institute of Technology GmbH. Giefinggasse 4, 1210 Vienna, Austria.}

\maketitle 

Future quantum networks distributing entanglement between distant quantum processors~\cite{Kimble2008, Wehner2018} hold the promise of enabling novel applications in communication, computing, sensing, and fundamental science~\cite{QuantumFingerprinting,fastQbyzantineagreement,baseline_telescopes,CHSH}.
Over the past decades a range of experiments on different qubit platforms have demonstrated the rudimentary capabilities of quantum networks at short distances including photon-mediated entanglement generation~\cite{Moehring2007, ritter_elementary_2012,Hofmann2012_heralded_entanglement_atoms, Bernien2013, stockill_phase-tuned_2017, stephenson_high-rate_2020, Krutyanskiy2023_ions_230m}. These short-range qubit networks are useful for testing of improved hardware~\cite{Drmota2023}, developing a quantum network control stack~\cite{Pompili_linklayer_2022} and for exploring quantum network protocols in a lab setting~\cite{Pompili_multinode_2021, Hermans_teleportation_2022, Drmota2024_VBQC_trapped_ions}. 

The next major challenge is to develop quantum network systems capable of generating, storing and processing quantum information on metropolitan scales. Such systems face several new requirements. First, the large physical distance, the consequential significant communication times and need for scalability demand that the network nodes operate fully independently. Second, as the optical fibers connecting nodes will extend for tens of kilometers, photon loss becomes a critical parameter that must be mitigated. Third, as advanced network applications require the heralded delivery of shared entangled states ready for further use, the qubit systems must be able to store quantum information for extended times and the network system must be capable of applying real-time feedback to the qubits upon successful entanglement generation.

Recent qubit experiments have shown promising progress towards the latter two criteria, including the integration with efficient quantum frequency converters~\cite{Bock2018, Krutyanskiy2019_light_matter_50km,Bock_Eich_Kucera_Kreis_Lenhard_Becher_Eschner_2018, Schaefer2023_twostage_conversion_SiV,  Tchebotareva2019,   Bersin2024}, demonstration of long coherence times on qubit systems that can be extended into multi-qubit registers~\cite{Bradley2019, Krutyanskiy2023_multimode_ionphoton_101km} and entanglement generation between nearby qubits via tens of kilometers of optical fiber~\cite{Leent_entanglement33km_2022, Knaut_SiV_35kmBoston_2023}. In parallel, experiments on ensemble-based quantum memories have pioneered significant advances on the first two criteria~\cite{LagoRivera2021, Luo_metropolitan_2mems_2022, Liu_metropolitan_network_3mems_2023, LagoRivera2023}.

\begin{figure*}
\centering
\includegraphics[width=1\textwidth]{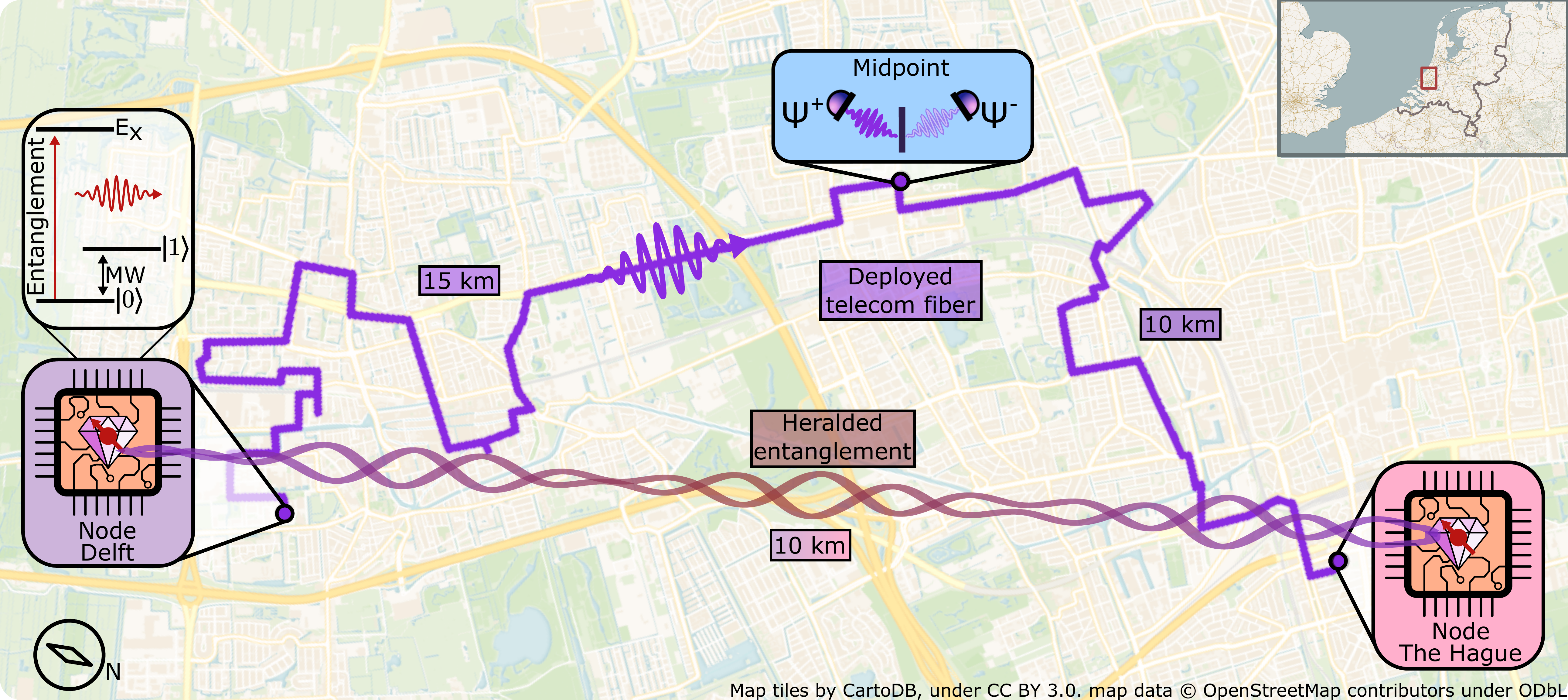}
\caption{\label{fig:fig1}\textbf{The metropolitan-scale quantum link.} Cartographic layout of the distant quantum link and the route of the deployed fiber bundle, with similar quantum processor nodes in Delft and The Hague. Fiber length between node Delft and midpoint is \SI{15}{\kilo\meter}, and between node The Hague and midpoint is \SI{10}{\kilo\meter}, with losses on the quantum channels at \SI{5.6}{\decibel} and \SI{5.2}{\decibel}, respectively. Inset to the quantum processor are the used qubit energy levels where the qubit is encoded in the electronic ground state addressable with microwave pulses (MW), and the spin-selective optical transition ($\lambda = \SI{637}{\nano\meter}$) is used for entanglement generation and state readout.}
\end{figure*}

Here, we report on the realization of a deployed quantum link between two solid-state qubit nodes separated by \SI{10}{\kilo\meter} matching all three criteria. 
The two network nodes are combined with a midpoint heralding station via \SI{25}{\kilo\meter} of deployed fiber, with all relevant classic and quantum signals propagating over the same fiber bundle in telecom bands (see Fig.~\ref{fig:fig1}). We implement an extensible architecture that enables the nodes to operate fully independently at large distance, mitigates the effects of photon loss on the entangling rate and allows for full heralding of entanglement generation. Furthermore, the network architecture features precise polarization and timing control as well as active stabilization of the relative optical phase between photons emitted from the nodes, enabling the use of the loss-resilient single-click protocol for efficient entanglement generation~\cite{Cabrillo_Cirac_García-Fernández_Zoller_1999, Bose1999}.
We benchmark the performance of the architecture by parameter monitoring and by generating entanglement in post-selection. Finally, we use the full network capabilities of heralding and real-time feedback to deliver entangled states shared between the nodes ready for further use. This demonstration establishes a critical capability for future applications and scaling and presents a key milestone towards large-scale quantum networking.
\vfill{}

\section{Deployed quantum network link architecture}
\begin{figure*}[ht]
\centering
\includegraphics[width=0.71\textwidth ]{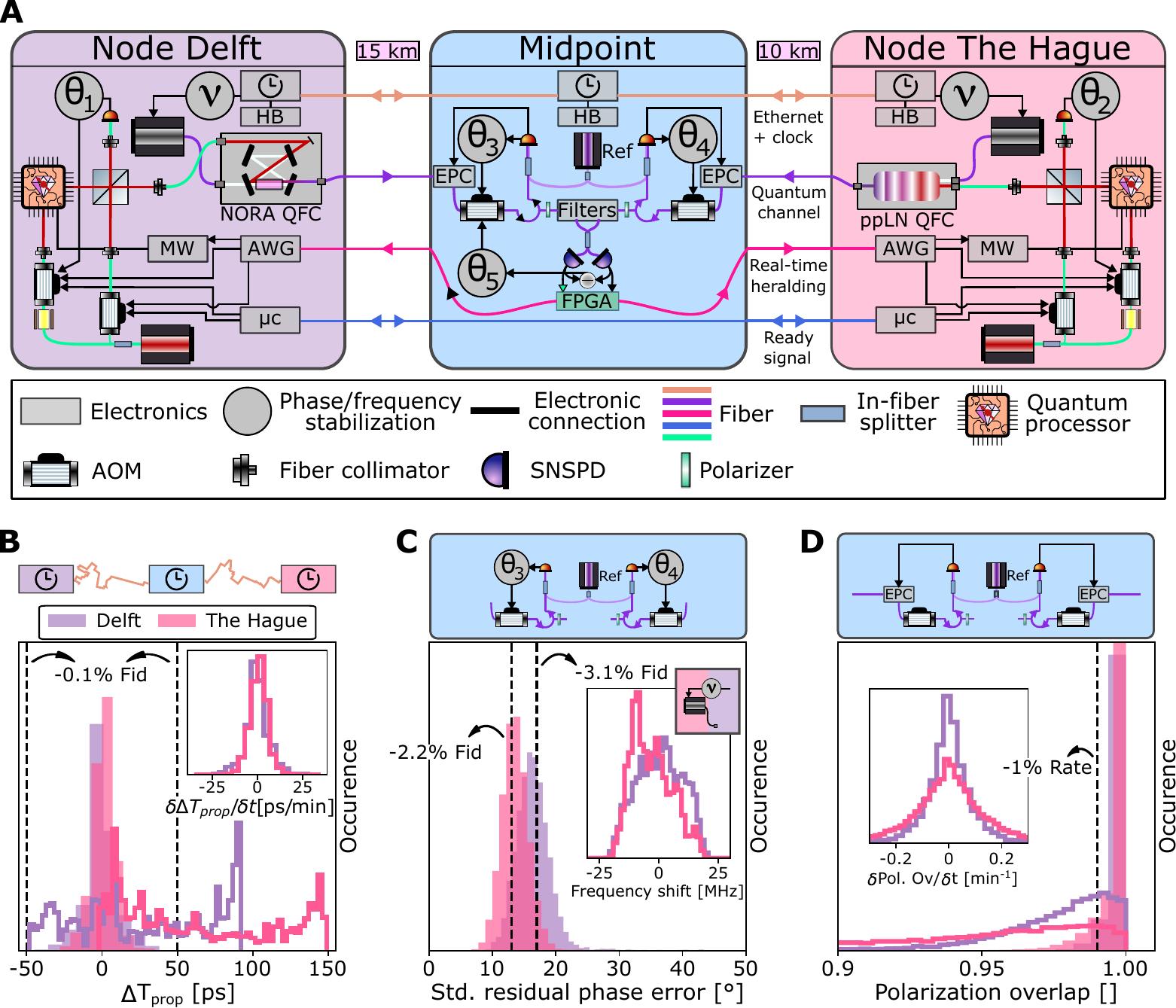} 
\caption{\label{fig:fig2}\textbf{Quantum node components and metropolitan-scale stabilization performance } \textbf{(A)}  Detailed components of the quantum nodes and fiber link connections. A micro-controller ($\mu$c) orchestrates the experiment, which together with an arbitrary waveform generator (AWG) shapes laser- and microwave (MW) pulses. Solid-state qubit entangled photon emission and stabilization light from each node is converted to the telecom L-band by the NORA (ppLN) based QFC in node Delft (node The Hague) and sent to a central midpoint. There, long distance qubit-qubit entanglement is heralded via single photon measurement (superconducting nanowire single photon detector (SNSPD), efficiency$\approx 60\%$, darkcount rate $\approx5s^{-1}$) with detection outcomes fed back in real-time. The stabilization light is used for phase locking at the nodes $\left(\theta_1,\theta_2\right)$, at the midpoint $\left(\theta_3, \theta_4, \theta_5\right)$ and for phase lock desaturation to the QFC pump lasers at the nodes $\left(\nu\right)$. The performance of stabilization over the deployed link over 24hrs is shown for \textbf{(B)} Time of arrival, \textbf{(C)} Phase and frequency and  \textbf{(D)} Polarization. Hardware providing active feedback (header) keeps these parameter that are drifting over time (line histogram) stable (shaded histograms) by enabling continuous feedback faster than the experienced drifts (insets). Vertical lines show the modeled impact on fidelity and rate.}
\end{figure*}

In order to meet the challenges of metropolitan-scale entanglement generation, we designed and implemented the control architecture depicted in Fig.~\ref{fig:fig2}A. Each node contains a CVD grown diamond chip hosting a Nitrogen-Vacancy (NV) center electronic spin qubit that can be faithfully initialized and read out by resonant laser light and controlled using microwave pulses. The NV center optical transition at~\SI{637}{\nano\meter} is used for generating qubit-photon entanglement.

Each node is equipped with a stand-alone quantum frequency converter (QFC) unit that converts the \SI{637}{\nano\meter} NV photons down to the telecom L-band at~\SI{1588}{\nano\meter} such that photon loss in the deployed fiber is minimized. The QFCs further serve as a tuning mechanism for compensating strain-induced offsets between the native emission frequencies typical for solid-state qubits. Through independent feedback on the frequency of the individual QFC pump lasers, we achieve conversion to a common target wavelength despite the few gigahertz difference in qubit emission frequencies~\cite{Stolk2022}. The QFC in Delft is based on a novel noise-reduced approach (NORA)~\cite{Geus_2023} that produces two orders of magnitude lower background counts than the  periodically-poled Lithium Niobate (ppLN) with integrated waveguide based QFC in The Hague~\cite{Dreau2018_conversion}.

To further mitigate photon loss, we employ the single-photon entangling protocol~\cite{Cabrillo_Cirac_García-Fernández_Zoller_1999, Bose1999, Hermans_singleclick_2023} for which the entangling rate favourably scales with the square root of the photon transmission probability across the entire link. In this protocol, each qubit is first prepared in an unbalanced superposition state $\ket{\psi} = \sqrt{\alpha}\ket{0}+\sqrt{1-\alpha}\ket{1}$. Application of an optical $\pi$-pulse resonant for qubit state $\ket{0}$ and subsequent spontaneous emission then results in qubit-photon entanglement, where the photonic qubit is encoded in the photon number state (0 or 1). Overlap of the photonic states at the beam splitter at the midpoint removes the which-path information, followed by single-photon detection by superconducting nanowire single photon detectors. Upon measurement of one photon after interference, entanglement of the qubit states is heralded to
$\ket{\Psi^{\pm}}=\left(|01\rangle \pm e^{i\theta}|10\rangle\right)/\sqrt{2}$ with maximum fidelity $1-\alpha$, where the $\pm$ sign is set by which output arm the photon was detected in. The entangled state phase $\theta$ is dependent on the optical phase difference between the photonic modes arriving at the interference beam splitter from Delft and The Hague. 

This entanglement generation critically relies on photon indistinguishability at the heralding station and therefore all degrees of freedom of the photons (frequency, arrival time, phase, polarization) must be actively controlled at the  metropolitan scale. A defining feature of our architecture is that stabilization laser light used for phase locking and polarization stabilization is time-multiplexed with the single-photon signals used for entanglement generation and sent over the same fiber from the nodes to the midpoint. To this end, we operate the link at a pre-defined heartbeat at \SI{100}{\kilo\hertz}, defining a common time division. During each heartbeat period of \SI{10}{\micro\second}, stabilization light is sent to the midpoint continuously, except for a \SI{2}{\micro\second} period where the entangling photonic states are sent out. This allows for near-continuous stabilization with high feedback bandwidth while performing entanglement generation.
Below we discuss the major sources of drift at metropolitan scale affecting the entangled state generation and our strategy for mitigating them.

First, length drift of the deployed fibers can result in reduced overlap of the photonic modes on the beam splitter. The timing of the optical $\pi$-pulse is locally controlled and disciplined by an optically linked distributed clock that doubles as Ethernet connection (White-Rabbit protocol~\cite{Dierikx_whiterabbitprotocol_2016}) over a dedicated fiber. Fig.~\ref{fig:fig2}B shows a histogram of measured offsets in time-of-arrival for the two deployed-fiber segments over 24 hours, with the inset displaying the drift speeds ($\sigma$ = \SI{5}{\pico\second\per\minute}). By using this data to compensate for the drift via timing adjustments of the control electronics every $\sim$15 minutes, the offset is kept below~\SI{50}{\pico\second}, much smaller than the photon $1/e$ decay time of~\SI{12}{\nano\second}. The resulting entangled state infidelity due to length drifts is below 0.1\%.

\begin{figure*}[ht]
\centering
\includegraphics[width = 0.69\textwidth]{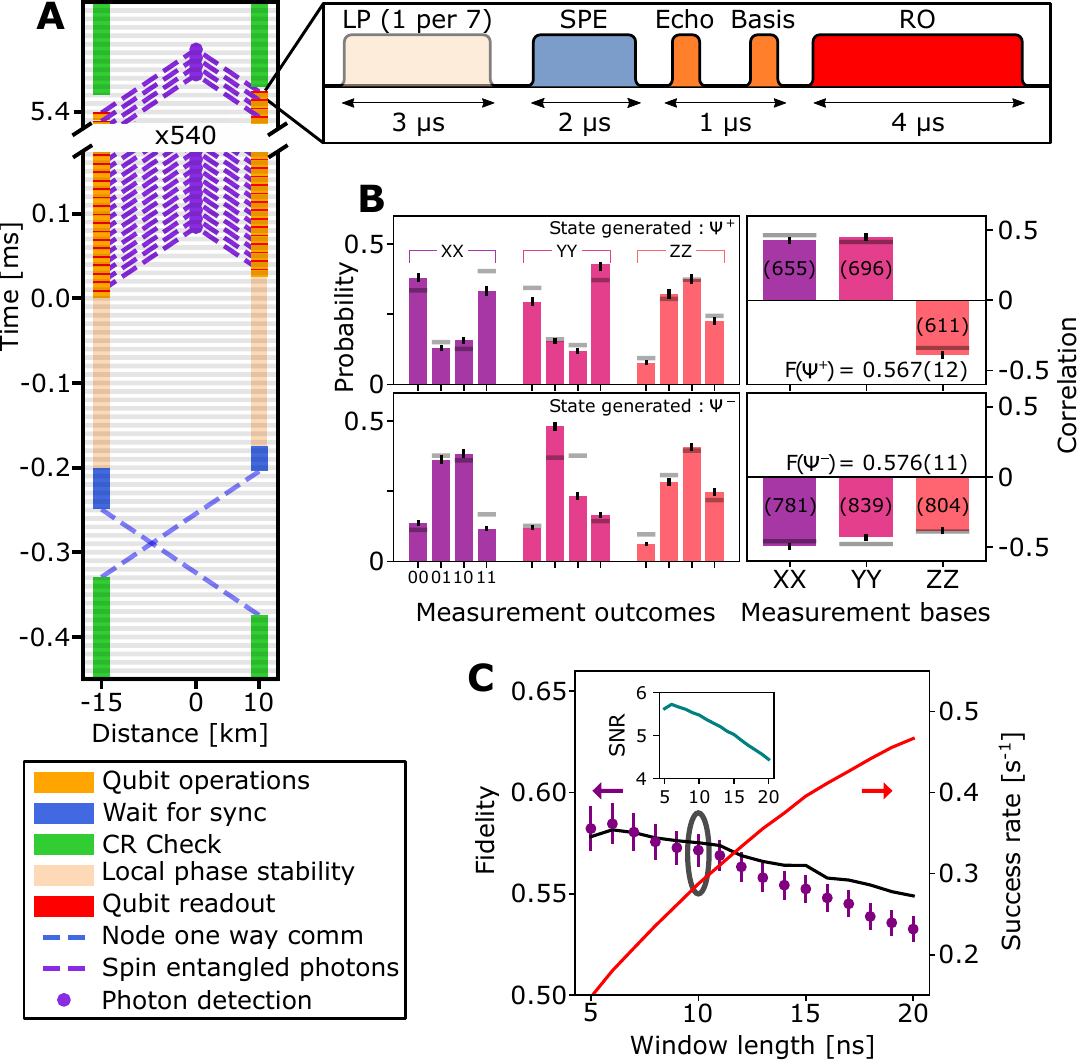}
\caption{\label{fig:fig3}\textbf{Post-selected entanglement over the deployed link} \textbf{(A)} Space-time diagram depicting the generation of entanglement in post-selection. Horizontal grey lines indicate the periodic heartbeat of \SI{100}{\kilo\hertz}. Local qubit control used to generate entanglement and perform state readout (pop-out) all fit within one heartbeat period of \SI{10}{\micro\second}. A local phase (LP) pulse is followed by spin-photon entanglement (SPE) generation, an echo and basis selection microwave pulse. Finally, the state is read out (RO).\textbf{(B)} Outcome of correlation measurements, with different detector signature for the top and bottom panels. We show the qubit-qubit readout outcomes per correlators(left), as well as the resulting values per correlator (right). The calculated state fidelity is given inside each figure. The number in parenthesis indicates the amount of events recorded for that correlator. Horizontal grey bars indicate the theoretical model. \textbf{(C)} Average state fidelity (left vertical axis) and entanglement generation rate (right vertical axis) for varying photon acceptance window length. Circling indicates the window used in (B), black solid line is a model with no free parameters (see Suppl. Mat.). Inset shows signal-to-noise ratio for the various window lengths. All measurement outcomes are corrected for tomography errors, errorbars are 1 standard deviation.}
\end{figure*}

Second, as the phase difference of the photonic modes interfering at the beam splitter is imprinted on the entangled state, this phase must be known in order for the generated entangled states to be useful. To achieve a known and constant optical phase setpoint, five individual phase locking loops are implemented across the total link, see Fig.~\ref{fig:fig2}A. At each node, a local phase lock is closed between reflection of the resonant excitation light off the qubit device surface and the stabilization laser light via the controllers $\theta_1$ and $\theta_2$, stabilizing the in-fiber and free-space excitation and collection optics. Phase noise from the long fiber and excitation laser is mitigated by the controllers $\theta_3$ and $\theta_4$ via interference of light from the midpoint reference telecom laser and frequency down-converted stabilization light at nanowatt levels. Importantly, analog phase feedback is performed locally at the midpoint directly on the incoming light yielding a high stabilization bandwidth exceeding \SI{200}{\kilo\hertz}.  Lastly, interference of telecom stabilization light from both nodes at the central beam splitter is measured by the single-photon detectors and input to controller $\theta_5$, closing the global phase lock between the nodes (see Suppl. Mat. for additional details). We note that the modularity of this architecture directly allows for the connection of multiple nodes to the same midpoint in a star topology, as the synchronization of all incoming signals to a central reference and relative phase stabilization between links can be performed using the control system at the midpoint.

Fig.~\ref{fig:fig2}C displays a histogram of the resulting phase errors during 24 hours of operation. Stable operation is achieved with a few-percent impact on the entangled state fidelity per connection. As this architecture yields full control over the phase difference at the beam splitter, the phase $\theta$ of the entangled state can be tuned on demand by adjusting the setpoint of the phase lock (see Suppl. Mat.). In order to maintain the phase lock under frequency drifts at the nodes, two individual feedback loops $\left(\nu\right)$ between the nodes and the midpoint adjust the frequency of the individual QFC pump lasers at an update rate of \SI{500}{\hertz}. Importantly, this desaturation allows for a large dynamic range of the phase feedback, as required to handle the observed frequency drift range of $>$\SI{50}{\mega\hertz}, see Fig.~\ref{fig:fig2}C inset. 

Third, polarization drifts, though significantly slower than phase drift, must also be mitigated. To this end, the stabilization light is additionally used for electronic polarization compensation (EPC) at the midpoint. The amplitude of the error signal input at $\theta_3$ and $\theta_4$ is dependent on polarization overlap with an in-fiber polarizer. We use this as input for a gradient ascent algorithm to feedback on the polarization of the incoming light at the midpoint. Data on the deployed link (inset Fig.~\ref{fig:fig2}D) shows that polarization drift occurs on second-timescales; our feedback at a few Hz bandwidth keeps the polarization aligned to within a few percent (Fig.~\ref{fig:fig2}D). Any remaining polarization mismatch is removed by the polarizers at the cost of a slightly reduced entanglement generation rate.

\section{Post-selected entanglement generation over a deployed link.}
We now turn to the performance of the deployed link in generating entanglement between the solid-state qubits at the remote nodes.
The proper functioning of all components of our system is first validated in a set of experiments with all devices of the link in a single lab in Delft, showing successful entanglement generation at state fidelities exceeding $0.6$ (see Fig. S3). After connecting and calibrating the equipment at the remote locations, we first focus on generating entanglement in post-selection. In this protocol the qubits are measured directly after generating spin-photon entanglement, and successful photon detection at the midpoint is used in post-processing to analyze entanglement generation. This scenario is compatible with quantum key distribution, but does not allow for more advanced protocols as the entangled state is not available for further use~\cite{Wehner2018}.

Our implementation of the post-selected entanglement generation is depicted in the space-time diagram of Fig.~\ref{fig:fig3}A, where each horizontal grey line depicts one \SI{10}{\micro\second} heartbeat period. Both nodes signal their start-of-experiment after passing their own charge-resonance check (CR Check) that ensures that the lasers are on resonance with the relevant optical transitions~\cite{Pompili_multinode_2021}. After communicating their readiness they resolve the earliest heartbeat to start attempting entanglement generation. In the first 20 heartbeat periods both nodes stabilize their local phase, followed by 540 rounds of entanglement generation, with one attempt per heartbeat period. Every seventh entanglement round also contains optical pulses for maintaining stability of the local phase. The operations performed at the nodes for each round are detailed in the pop-out of Fig.~\ref{fig:fig3}A. The sequence returns to the CR Check after completing the preset number of rounds.

We characterize the generated non-local states by measuring qubit-qubit correlations in different readout bases. In Fig.~\ref{fig:fig3}B we plot the outcomes for the three basis settings split out per detector, showing the expected (anti-)correlations. Combining the outcome probabilities we calculate the correlators $\langle ZZ\rangle$, $\langle XX\rangle$ and $\langle YY\rangle$. Note that as the two detectors herald different Bell states ($\Psi^{+}$ vs $\Psi^{-}$), the corresponding $\langle XX\rangle$ and $\langle YY\rangle$ correlations have opposite sign. We also plot the values predicted by our detailed model without any free parameters (grey lines, see Suppl. Mat.), and observe good agreement with the data. The asymmetry in the amount of events is caused by a difference in quantum efficiency between the two single-photon detectors. We find that the measured fidelities $F(\ket{\Psi^{\pm}}) = \frac{1}{4}*\left(1-\langle ZZ\rangle\pm\langle XX\rangle \pm \langle YY\rangle \right)$ with respect to the ideal Bell states are significantly above $0.5$, proving the generation of post-selected two-qubit entanglement (Fig.~\ref{fig:fig3}B).

In the above analysis, we included photon emission up to \SI{10}{\nano\second} after the optical $\pi$-pulse. By varying the analysis window of the photon detection, we can explore the trade-off between rate and fidelity (Fig.~\ref{fig:fig3}C). We find that the entangled state fidelity slowly decreases with increasing window length, following the signal-to-noise ratio of the photon detection at the midpoint (see inset to Fig.~\ref{fig:fig3}C). At the same time, as more photon detection events are accepted with increasing window length, the success rate increases. The achieved entanglement generation rate reaches \SI{0.48}{\hertz} (success probability per attempt of $7.2\cdot10^{-6}$) for a \SI{20}{\nano\second} window.

\section{Fully Heralded entanglement generation over a deployed quantum link}
In a final demonstration that highlights the capabilities of the deployed platform, we generate fully heralded qubit-qubit entanglement. In contrast to the post-selected entanglement generation described above, ``live'' entangled states are now delivered to the nodes that can be further used for quantum information tasks. Such live entanglement delivery is a fundamental requirement for many future applications of long-range entangled states~\cite{Wehner2018}. 

\begin{figure*}
\centering
\includegraphics[width = 0.8\textwidth]{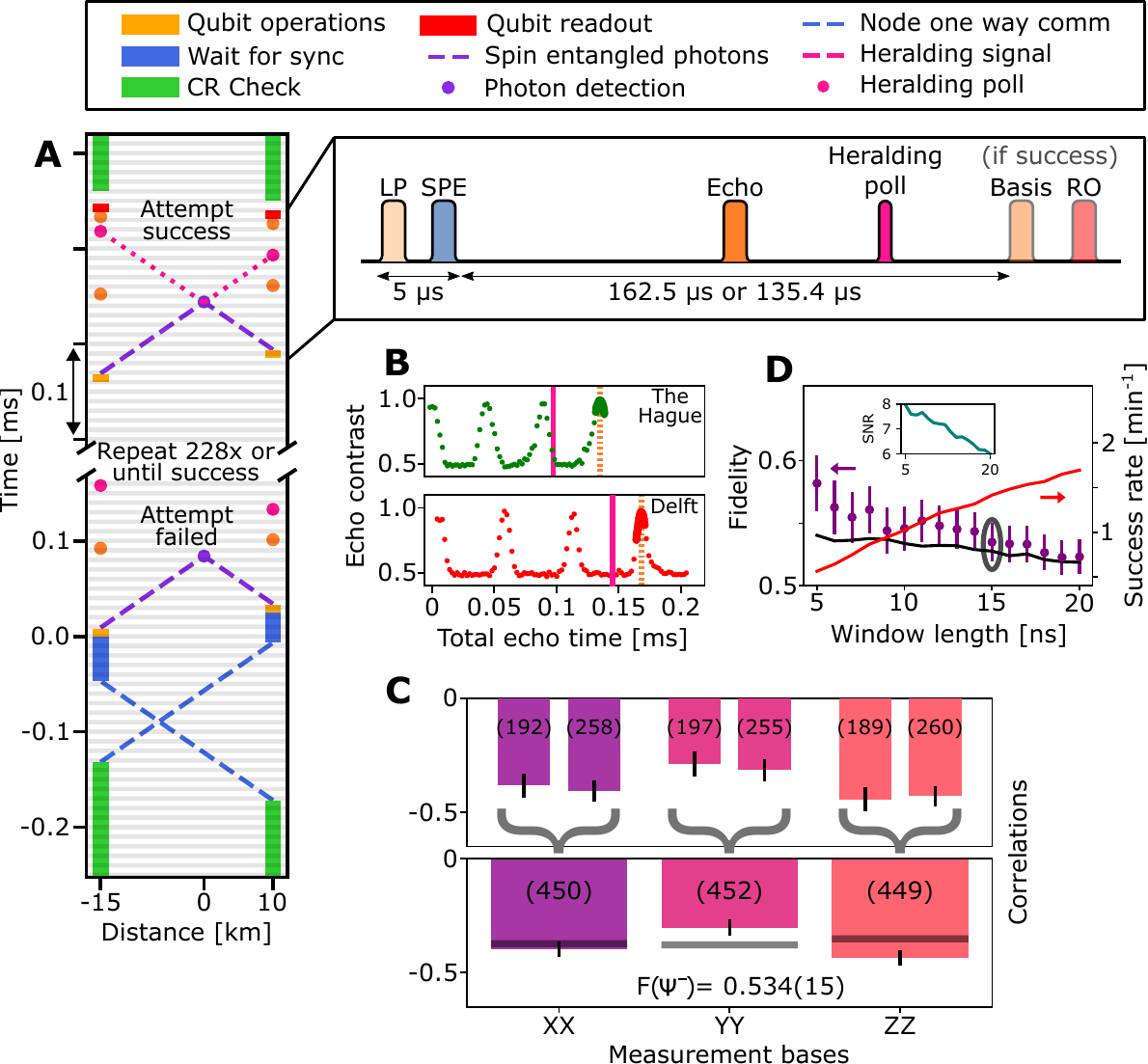}
\caption{\label{fig:fig4}\textbf{Fully heralded entanglement over the deployed link} \textbf{(A)} Space-time diagram of fully heralded entanglement generation. An attempt is successful upon registering a heralding signal at the polling time, after which a feed-forward is applied on the qubit and readout is performed. Absence of a heralding signal communicates a failed attempt, where we retry for a maximum of 228 attempts or until success. Pop-out depicts the local qubit control, basis selection and readout pulses. The time between spin-photon entanglement (SPE), heralding poll and basis selection is node-dependent. \textbf{(B)} Hahn-echo experiment on the communication qubit, showing the revivals of the coherence~\cite{Ryan2010}. Solid vertical line indicates the heralding poll, the dotted line the time of the basis selection. All times are with respect to echo sequence start. \textbf{(C)} Correlation measurement for full heralding, showing both detector outcomes delivering the same $\Psi^{-}$ state. Upper (lower) plot shows events per detector (combined). Bars indicate data, the number in parenthesis indicate the amount of events. Horizontal lines indicate the theoretical model. \textbf{(D)} Average state fidelity (left axis) and entanglement generation rate (right axis) for varying photon acceptance window length. Circling indicates the window used in (C), black line is a model with no free parameters (see Suppl. Mat.). Inset shows signal-to-noise ratio for respective window lengths. Measurement outcomes are corrected for tomography errors, errorbars are 1 st.dev. }
\end{figure*}

We emphasize that this protocol requires that all relevant heralding signals (including which detector clicked) are processed at the node before the entanglement delivery is completed. To this end, we employ the experimental sequence depicted in Fig.~\ref{fig:fig4}A. To preserve the qubit states with high fidelity while waiting for the heralding signals to return and be processed, the refocusing echo pulse is applied to the qubits halfway the sequence to dynamically decouple them from spin bath noise in their solid-state environment. Fig.~\ref{fig:fig4}B shows the resulting qubit preservation as a function of time. While the qubits are protected at the nodes, the photons travel to the midpoint in about \SI{52}{\micro\second} and \SI{73}{\micro\second} from The Hague and Delft, respectively. An FPGA at the midpoint processes the output of the single-photon detectors, establishing whether a photon was detected in a predetermined time window and in which detector. The electronic output of the FPGA is optically communicated to the nodes, taking another \SI{52}{\micro\second} (\SI{73}{\micro\second}). There, the signal is detected and processed live to choose the next action. The time at which this final processing is completed per node is indicated by the solid pink lines in Fig.~\ref{fig:fig4}B.

We choose to use the first echo revival (orange dotted vertical line in Fig.~\ref{fig:fig4}B) after the expected arrival time of the heralding signal to complete the delivery of the entangled state. For these setpoints, a detailed measurement of the echo contrast shows that the wait times introduce a $1.3$ ($3.2$) percent reduction in coherence for node Delft (The Hague). We note that using more advanced dynamical decoupling sequences the NV qubit coherence time could be extended towards a second~\cite{Abobeih_2018}.

The entanglement generation runs are automatically repeated by the nodes until a successful heralding signal is received from the midpoint. Once such a signal is received, the system jumps to a different control sequence in which a basis selection gate is applied to each qubit followed by single-shot qubit readout. This control sequence incorporates the which-detector information communicated from the midpoint in real time. We exploit this full heralding capability to apply a phase flip conditioned on which detector clicked, thereby delivering the same Bell state $\Psi^{-}  = \frac{|01\rangle - |10\rangle}{\sqrt{2}}$ for each of the two possible heralding signals.

The measured correlations per readout basis are shown in Fig.~\ref{fig:fig4}C, showing the expected anti-correlated outcomes for all three bases. The top bars show the outcomes divided per detector, displaying now the same $\Psi^{-}$ thus showing the successful operation of the real-time feedback. We find that the delivered entangled states have a fidelity of $0.534(15)$. This result establishes the first demonstration of heralded qubit-qubit entanglement at metropolitan scale, with all the heralding signals processed in real-time and the entangled states delivered ready for further use. 

Fig.~\ref{fig:fig4}D displays the rate-fidelity trade-off in analogy to Fig.~\ref{fig:fig3}C, showing a similar trend. The reduction in rate ($\SI{0.022}{\per\second}=\SI{1.3}{\per\minute}$ at the pre-defined \SI{15}{\nano\second} window length) compared to the post-selected case is mainly due to the added communication delay needed for the heralding signal to travel to the nodes, making each attempt a factor of $\approx 20$ slower. The observed fidelity vs. window size is well captured by our model without any free parameters, with the reduction in fidelity compared to the post-selected case mainly coming from the additional decoherence and a reduced phase stability (black line, see Suppl. Mat.). The improved SNR (Fig.~\ref{fig:fig4}D inset) compared to Fig.~\ref{fig:fig3}C is due to an improved trade-off between detection efficiency and dark counts following an optimization of the single-photon detector bias currents. 

\section{Conclusion and outlook}
We have realized a deployed quantum link and demonstrated heralded entanglement delivery between solid-state qubits separated at metropolitan scale. The architecture and methods presented here are directly applicable to other qubit platforms~\cite{Lodahl_2017, Bradac_2019_nanophotonics_groupIV,Son2020_SiC_overview,Ruf_Wan_Choi_Englund_Hanson_2021, Leent_entanglement33km_2022, Krutyanskiy2023_ions_230m} that can employ photon-interference to generate remote entanglement and frequency conversion to minimize photon losses. Additionally, the ability to phase-lock remote signals without the need for ultra-stable reference cavities can be of use for ensemble-based quantum memories~\cite{Luo_metropolitan_2mems_2022, LagoRivera2021,Rakonjac_Grandi_Wengerowsky_Lago-Rivera_Appas_Riedmatten_2023, Liu_metropolitan_network_3mems_2023}.

This work can benefit from future developments in the following ways. Real-time correction of false heralds~\cite{ Hermans_teleportation_2022} can be realized by using detection events of phonon side-band photons on the nodes, upon which the fidelity of the delivered entangled state improves (see Suppl. Mat.).  Near-term developments can substantially improve the signal-to-noise ratio, which is currently limiting the entangled state fidelity (about 30\% contribution). For instance, the signal can be significantly boosted by embedding the NV-centre in an open micro-cavity~\cite{Ruf_Weaver_van_Dam_Hanson_2021, Riedel2017_enhacement_NV} or by using different color centers that exhibit a more efficient spin-photon interface~\cite{Sipahigil2016, Bersin2024,Knaut_SiV_35kmBoston_2023, SnVwaveguidesPasini}. As a quantitative example, replacing the NV center with a diamond SnV center (which has a 16 times higher probability of coherent photon emission), employing NORA QFCs at all nodes~\cite{Geus_2023} (here used only in the Delft node), and fixing a known imperfection in the local phase stabilization would already increase the heralded state fidelity to above 80\%. Furthermore, improving coherence protection using XY4 sequences and systematic reduction of the remaining small error sources could lift the fidelity beyond 90\%  (see Table S4). The extensible nature of our architecture opens the door to connecting more than two qubit nodes to a midpoint, which, in combination with using local memory qubits~\cite{Bradley2019, stas2022_error_detection, Drmota2023,Krutyanskiy2023_repeater}, would enable the exploration of more advanced protocols on a metropolitan-scale network~\cite{Kalb_Hanson_2017_distillation, Hermans_teleportation_2022,Drmota2024_VBQC_trapped_ions}, as well testing quantum control stacks~\cite{Pompili_linklayer_2022} on a distributed multi-node quantum network.

\section*{Acknowledgements}
We thank Christine Satter, Castro do Amaral, Eftychia Tsapanou-Katranara and Jacob Dalle for help with characterizing and fabricating the diamond device, Leon van Dooren for help in developing the local phase stabilization implementation,  Jaco Morits for system engineering support, Theo Lodewijkx for managing the engineering tasks in the project, Ingrid Romijn for program management support and Stephanie Wehner for project development in the early stages of the project, Otmar Ubbens for support on various stabilization tasks, Sidney Cadot, Henri Ervasti, Ravi Budhrani for software development and support, Raymond Vermeulen and Raymond Schouten for development of custom electronics, Siebe Visser and Vinod Narain for technical assistance during the moving of the equipment, Laurens Feije for assistance with the move of the midpoint, Marco Gorter and Nico Coesel for providing hardware support with the OPNT time synchronization devices, and Paul Slootmaker and Ed van de Bovenkamp for their organizational and on-site support at the KPN locations.
\textbf{Funding:} We acknowledge funding from the Dutch Research Council (NWO) through the project “QuTech Part II Applied-oriented research” (project number 601.QT.001), the Spinoza prize 2019 (project number SPI 63-264), and the Zwaartekracht program Quantum Software Consortium (project no. 024.003.037/3368). We further acknowledge financial support from the EU Flagship on Quantum Technologies through the project Quantum Internet Alliance (EU Horizon 2020, grant agreement no. 820445) and from the Dutch Ministry of Economic Affairs and Climate Policy (EZK), as part of the Quantum Delta NL programme, and Holland High Tech through the TKI HTSM (20.0052 PPS) funds. Finally, we acknowledge funds from Fraunhofer-Gesellschaft zur Förderung der angewandten Forschung e.V. and Fraunhofer Institute for Laser Technology ILT through the ICON program. 
\textbf{Author contributions:}
A.J.S., K.L.vd.E., E.v.Z. and R.Hanson devised the experiment. 
R.W.H., W.D.K., A.J.H.M., R.Hagen, J.J.B.B. tested and built experimental hardware, prepared the remote locations and supervised moving. 
J.v.R., P.B. and I.t.R-D. developed the software for remote control and monitoring of equipment, and heralding with input from A.J.S. and K.L.vd.E. 
M.M. and A.M.E. grew and prepared the diamond device substrates. 
R.V. fabricated one of the diamond devices.
J.F.G, F.E., B.J. and C.H. designed, built and tested the NORA QFC apparatus.
C.T., J.S., S.R. designed, built and tested the low phase noise excitation lasers.
A.S. and K.L.vd.E. wrote the measurement code, carried out the experiments and analyzed the data.
A.J.S. and M.-C.S. developed supporting simulations. 
A.J.S., K.L.vd.E. and R.Hanson wrote the main manuscript. 
A.J.S. and K.L.vd.E. wrote the supplementary information. 
R.Hanson conceived the project and supervised the experiments.
All authors commented on the manuscript.
\textbf{Competing interests:} The authors declare no competing interests.
\textbf{Data and materials availability:} The data supporting this manuscript is available at 4TU.ResearchData~\cite{manuscript_data}.

\newpage

\bibliography{main}
\bibliographystyle{my_plain}

\onecolumngrid
\clearpage

\begin{center}
\textbf{\Large Supplementary Materials}
\end{center}

\makeatletter
\renewcommand{\theequation}{S\arabic{equation}}
\renewcommand{\thefigure}{S\arabic{figure}}
\renewcommand{\thetable}{S\arabic{table}}
\renewcommand{\thesection}{S\arabic{section}}
\makeatother
\setcounter{equation}{0}
\setcounter{figure}{0}
\setcounter{table}{0}
\setcounter{section}{0}

\section{Experimental setup}
The experimental set-ups used in this work are built on top of the hardware described in Ref.~\cite{Stolk2022} and they use software based on the Quantum Measurement Infrastructure, a Python 3 framework for controlling laboratory equipment~\cite{raa_qmi_2023}. In this section we will discuss the hardware changes made that enabled compatibility with metropolitan-scale entanglement generation.

\subsection{Quantum Node}
We exchanged the diamond quantum device of Node The Hague for a newly fabricated one. All parameters of the substrate (Element Six) and device, such as purity, crystal orientation and carbon isotopes are the same as previously reported.
To allow for phase stability between physically separated setups, the TOPTICA DLC DL pro \SI{637}{\nano\meter} was exchanged for an upgraded prototype employing optical feedback from an additional cavity to reduce the phase noise to $<\SI{40}{\milli\radian}$ integrated from \SI{100}{\kilo\hertz} to \SI{100}{\mega\hertz}.  
\subsection{Quantum Frequency Converters}
As mentioned in the main text, we have replaced one of the in-house built QFC modules with the NORA QFC described in~\cite{Geus_2023}. This design mitigates the amount of noise photons generated by the frequency conversion due to imperfections in the waveguide and poling period of periodically poled Lithium Niobate (ppLN) crystals. Briefly, the novel approach consists of identifying the right bulk material that can be critically phase-matched to perform efficient conversion, and embedding the crystal in a single-resonant cavity to enhance the effective pump power in the conversion process. We compare the NORA QFC to the ppLN QFC in Table~\ref{supptable:qfc_comparison}, and refer to~\cite{Geus_2023} for more information.

\begin{table}[h]
\begin{center}
\begin{tabular}{| c |M{1.3cm}| M{2cm} | M{1.5cm} | M{2.2cm} | M{2.2cm} |}
\hline
 QFC & Node & Non-linear medium & Efficiency (front to end) & Noise $\left[\SI{}{\per\second\per\giga\hertz}\right]$ & Eval. at filter bandwidth\\
 \hline\hline
 QuTech~\cite{Tchebotareva2019,Stolk2022} & The Hague & ppLN waveguide & $50\%$ & $2104$ & \SI{50}{\mega\hertz} \\
 \hline
 Fraunhofer ILT~\cite{Geus_2023}& Delft & bulk KTA & $48\%$ & $19$ & \SI{374}{\giga\hertz}\\
 \hline
\end{tabular}
\caption{Comparison of the QFCs used in the experiment. }
 \label{supptable:qfc_comparison}
\end{center}
\end{table}

\subsection{Phase stability on the nodes}
On both nodes we added the components to define the local interferometers that lock the excitation laser path to the stabilization path of the respective set-ups. We use a combination of up- and down-shift Acousto-Optical Modulators (AOM, Gooch and Housego Fiber-Q), already used for amplitude modulation, to define a \SI{400}{\mega\hertz} frequency offset. The excitation light reflected off the diamond surface is separated from the zero-phonon-line (ZPL) path by a polarizing beam splitter (PBS), where it spatially overlaps with the stabilization light. We guide the light to a set of birefringent crystals that maximize the complex overlap between the orthogonally polarized reflected beam and the stabilization light. A polarizer then selects a common polarization such that both beams can interfere in intensity, after which we focus the light into a multi-mode fiber (core diameter $\approx 100um$) and measure the interference beat signal using a fast avalanche photodiode (Menlo Systems APD210). After amplification and filtering, we extract the phase of the interference by mixing it with an electronic reference signal generated by a stable RF source (Anapico APSIN6010). The feedback signal generated is passed through a Track-And-Hold amplifier (Texas Instruments OPA1S2384) and input to the Frequency Modulation input of the Excitation light AOM RF-driver (TimeBase DIM3000), which closes the feedback loop. Because the frequency modulation has integral feedback on the phase that can not be switched on/off fast, any bias/offset in the error-signal reduces the free evolution time during which the local phase remains stable. In a future implementation, this actuator will be replaced by a linear phase shifter, and an improvement of the local phase-stability can be realized.

\subsection{Midpoint}
The previous design of the midpoint used for two-photon quantum interference~\cite{Stolk2022} suffered from an increased noise floor due to the presence of bright laser pulses incident on the nanowires every \SI{200}{\micro\second}. To reach an adequate feedback bandwidth for phase-locking of our excitation lasers to the telecom reference, we have to reduce the period of bright pulses to \SI{10}{\micro\second}, making this even more challenging. We have made the following improvements to allow for polarization and fast phase feedback, to reduce crosstalk between optical channels and lower the background counts in the SNSPDs:
\begin{itemize}
    \item Split the high-power reference light into three paths, two used to generate an optical beat to lock the stabilization light coming from each node to the reference laser, and one used to generate an error signal to stabilize the ultra-narrow Fibre-Bragg grating filters.
    \item Added two low-loss Variable Optical Attenuators (VOAs, Boston Applied Technologies Nanona VOA) that shield the detectors from bright pulses coming from the nodes.
    \item Added two Electronic Polarization Controller (OZOptics EPC-400-11-1300/1550-9/125-S-3A3A-1-1 ) to allow for full control over the polarization state to compensate for fiber drifts.
    \item Added two AOMs to have fast and integral control of the phase of the incoming light, used for phase stabilization.
    \item Removed one circulator from the design and splicing multiple connectors where the single-photons pass through to reduce single-photon loss
\end{itemize}

All the error signals for the phase- and polarization stabilization, and the ultra-narrow fiber Bragg-grating (FBG) for one node are generated on the same balanced photodiode (Thorlabs PDB480C-AC), and subsequently extracted by a combination of power-splitter, bandpass filters and amplifiers.

\subsection{Heralding}
To allow for low-latency feedback of photon detections at the midpoint, the midpoint was upgraded with an FPGA development board (Digilent Arty-A7) with custom-built PCBs converting the external pin I/O to SMA connectors with the possibility of \SI{50}{\ohm} matching. Single photon signals were input on a custom pulse-stretcher, cleaning up the electrical pulse from the SNSPD to a block-like pulse shape of around \SI{1}{\micro\second} and input on the I/O of the FPGA. The complete I/O of the FPGA is shown in Fig.~\ref{suppfig:fig9}. The acceptance window of sending out a heralding signal is predetermined at the time within a heartbeat period where we expect spin-photon entangled photons to arrive. The exact timing of the acceptance window is derived from the centralized clock and heartbeat signal available at the midpoint. To be able to signal which detector clicked on a photon count, we use two digital pulses to herald, one for the success, and one for which detector clicked. Since the delay in time-of-flight between the nodes is different, the digital pulses need to be aligned individually to arrive at the nodes at the same time within one heartbeat. To clarify, this does not mean they arrived at the same global time at the nodes, but that the digital pulses arrive at the same time modulo the heartbeat (\SI{10}{\micro\second}) at the nodes.
Lastly, a local FPGA code loading server was employed that could upload compiled-VHDL that was either manually or automatically generated remotely. A change in timing of sending out pulses could be updated on the underlying VHDL code automatically based on a custom template code. This allowed fast multi-core compilation locally and remote upload to significantly reduce debugging time when determining the required exact timing of the heralding pulses.

\section{Phase stabilization in the midpoint}
\begin{figure}[ht]
\includegraphics[width=\textwidth]{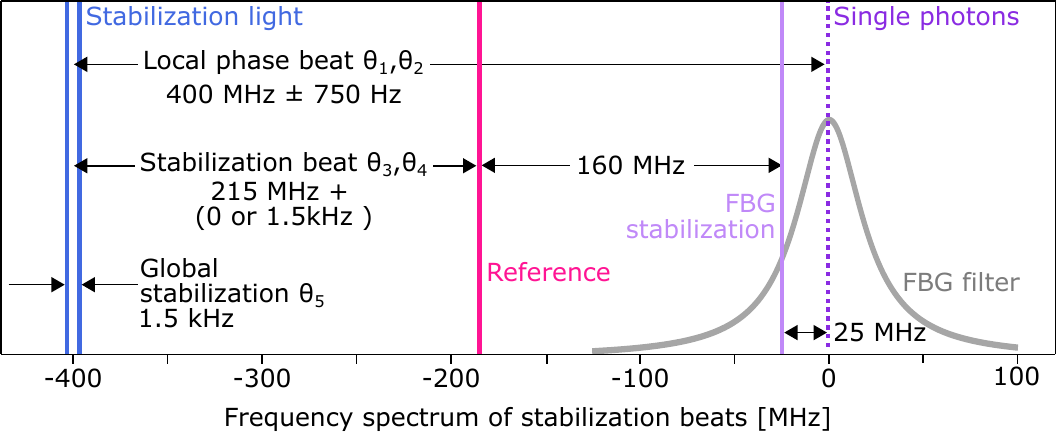}
\caption{\label{suppfig:fig1}Frequency of relevant optical fields for the phase lock in the midpoint. The grey Lorentzian shape is the ultra-narrow fibre-bragg grating filter (FBG). It is stabilized using a \SI{160}{\mega\hertz} beat generated by the reference laser and a frequency-shifted tap-off that is back-propagated through the flank of the FBG (\SI{25}{\mega\hertz} from the transmission peak). The stabilization light is located far from the filter resonance, detuned by \SI{400}{\mega\hertz} from the filter peak and excitation laser/single-photons, which is locked by the local phase lock controllers $\theta_1$ and $\theta_2$. This stabilization light is reflected off the FBG and interferes with the reference laser generating a $\approx\SI{215}{\mega\hertz}$ beat, which is stabilized by the fast phase lock controller $\theta_3$ and $\theta_4$. The slight offset between the nodes of \SI{1.5}{\kilo\hertz} is measured by the SNSPDs and input to controller $\theta_5$, to close the phase lock between Delft and The Hague.}
\end{figure}

We stabilize the incoming light from each node to the same telecom reference (fast stabilization). The relative optical phase between the nodes is stabilized using interference at the central beam splitter in the midpoint, and measured with the SNSPDs (global stabilization). The optical fields and respective frequencies are shown in Fig.\ref{suppfig:fig1}, and the distribution of the fields in time shown in Fig.~\ref{suppfig:fig4} and~\ref{suppfig:fig5}.
\subsection{Fast stabilization}
Part of the stabilization light used in the local phase lock also propagates via the QFC, through the same deployed fiber, to the midpoint. There it is split off in frequency from the single-photons via the FBG, subsequently interfered with a telecom reference laser (NKT Adjustik fiber laser), where a heterodyne interference beat is measured by the balanced photodiode. A phase-frequency detector (Analog Devices HMC3716LP4E) processes this signal by comparing it to a reference signal (Wieserlabs FlexDDS-NG DUAL). Its output serves as the error-signal for a high-speed servo controller (Newport LB1005-S) performing PID control. This controller provides feedback on the AOM to stabilize the incoming light, synchronizing it with the reference at a bandwidth of over \SI{200}{\kilo\hertz}. To stay within the low-loss performance of the AOM close to the central frequency, we offload the accumulation of phase error (frequency drifts) to the QFCs at the nodes. This can be done much slower (\SI{500}{\hertz}) and makes the feedback only limited by the frequency range of the QFCs ($\gg$\SI{1}{\giga\hertz}). The fast feedback removes all high-frequency noise from the incoming light, which contains the excitation laser line-width, as well as phase noise introduced by expansion/contraction and vibrations of the deployed fiber.
\subsection{Global stabilization}
To further compensate phase drift that occurs in the midpoint and to set the relative phase between the incoming optical modes, we employ a control loop based on the stabilization light from both nodes that interferes at the central beam splitter. This light leaks through the FBG, and is of sufficient low power to not blind the SNSPDs. The voltage pulses from the SNSPDs are amplified, and a difference amplifier (Krohn-Hite Model 7000) generates a heterodyne beat by subtracting the count-rates from each detector. This beat is compared to a reference signal (AimTTi) by mixing it and low-pass filtering (\SI{150}{\hertz}), generating an error-signal. This is again input to a servo controller (Newport LB1005-S), of which the output is used to modulate the phase of the reference of one of the fast stabilization controllers.
\subsection{Verifying phase stability}
We can verify the performance of the full optical phase stabilization by interfering the two excitation lasers from the nodes at the central beam splitter in the midpoint, using the time-resolved counts in the SNSPDs to generate the error signal. By changing the phase of the reference signal, we can measure a full visibility fringe of the interference. By correcting the measured interference for the imbalance in photon flux from each node, we can calculate the interference contrast, see Fig.~S2B. A more in-depth description and technical analysis of the total phase stabilization will be given in a separate manuscript~\cite{StolkPhase2024}.

\section{Overview communication times and decisions}
In this section we discuss the different optical and electrical pulses or feedback signals that are needed to control this distributed experiment. We will first discuss the local nodes, then which signals are necessary to be timed at the midpoint for single photon processing and phase stability. Lastly, we will discuss how we align the arrival time of the photonic states at the midpoint telecom detectors.
\subsection{Pulses and timing at node}
Since the central clock in the midpoint is distributed to the nodes by the White Rabbit enabled switches, it provides a common reference frame to align the time-of-arrival for each of the entanglement generation attempts, see Section~\ref{section:tof_alignment}.
A full timing overview of a single heartbeat length when we perform a full entanglement attempt without heralding is shown in Fig.~S4A and with heralding in Fig.~\ref{suppfig:fig4}B. The entanglement generation attempt with heralding contains exactly the same pulses as in Fig.~S4A, except for that the time between the spin superposition state generation and basis-selection pulse is at a revival of the spin coherence, as previously shown in the pop-out of Fig.~4A of the main text.
\subsection{Pulses and timing at midpoint}
The laser, electrical and feedback pulses used at the midpoint by design operate agnostically of the experiment that is being run. For about $\sim$\SI{8}{\micro\second} we can perform phase, frequency and FBG stability actions using the interference from the reference light and the incoming stabilization light from the node (see Fig.~S5). In the remaining $\sim$\SI{2}{\micro\second} the local reference light is shut off, the phase Track-and-Hold amplifier is set to `hold' and the VOAs opened fully to allow for single photons to pass through to the SNSPDs. 

\subsection{Time-of-arrival alignment}
\label{section:tof_alignment}
One of the requirements for indistinguishability of the photons is the time aligned photonic mode overlap at the beam splitter in the midpoint. The timing of the emission of the photonic state determines, together with the time of flight (ToF), the time at which the photonic state arrives from the node at the midpoint. This is especially crucial in the metropolitan distance case, where fiber distance varies with tens of picoseconds per hour (see Table~\ref{table:fibers}). Since we have unbalanced fiber connection lengths between the nodes-midpoint, Node Delft will be allowed to send its photon as soon as possible. Node The Hague has to hold its pulse sequence start for 3 heartbeats. Both nodes require sub-heartbeat timing alignment to align the arrival time of the solid-state entangled photons in the last $\sim$\SI{2}{\micro\second} of the midpoint scheme (Fig.~5). This last alignment step is achieved with the arbitrary waveform generator (AWG, Zurich Instruments HDAWG) sequencer and the built in skewing parameter of the AWG outputs.

Due to the constantly changing ToF, the alignment in time is performed in two steps. A first time alignment is required to be performed manually where laser pulse timing is recorded at the midpoint timetagger (PicoQuant Multiharp, \SI{80}{\pico\second} resolution), after which the time difference between arrival times of the pulses from the two nodes can be calculated and adjusted. This value is recorded as calibration parameter. Secondly, the reported ToF from the time synchronization system is used to automatically adjust the skewing parameter every $\sim$15 minutes with the difference in fiber ToF with respect to the initial calibration. 

\section{Other midpoint stabilization}
\subsection{Polarization stabilization}
We continuously stabilize the polarization of the incoming stabilization light using the same beat generated on the balanced photodiode as mentioned in the phase stability section (Section S2) . Because of our in-fiber polarizer, the amplitude of the beat will depend on the overlap of the incoming light with the polarizer. Maximizing this amplitude will guarantee the same polarization between the nodes, as needed for interference. Using a phase-gain detector (Analog Devices EVAL-AD8302) we can monitor the amplitude of the beat by comparing it to an in-phase RF tone (generated by Wieserlabs FlexDDS-NG DUAL) over multiple orders of magnitude. We sample the output of this phase-gain detector with an Analog Discovery 2 (Digilent), and process the error signal in a python process running a Gradient Ascent algorithm. This in turn sends commands to the EPC that adapts the polarization state and completes the control loop. The whole process can be monitored and adjusted via a Graphical User Interface.
\subsection{Ultra-narrow FBG stabilization}
We adapted the control loop for the temperature stabilization of the ultra-narrow FBGs (UNF) to allow for a significant reduction of the optical power reflected back to the SNSPDs. We achieved this by using a heterodyne beat measurement that is sensitive to the transmission through the UNF, instead of a simple power measurement at DC. This allows for multiple orders higher sensitivity, and allows us to stabilize the filters with only $\approx$ \SI{10}{\femto\watt} of light traveling through the filter in the backwards direction.
\section{Deployed fiber characterization and stability}
We use a fiber bundle with four fibers to perform all the communication between the locations. These are provided to us by KPN, a major Dutch telecom provider, and are part of their telecom infrastructure. The reality of using deployed fibers means that it is built up of sections of continuous fiber, which are spliced or connected together at various locations along the path. Therefore the losses per kilometer of deployed fiber are significantly higher than the ideal losses of $\approx$\SI{0.2}{\decibel\per\kilo\meter} @ \SI{1580}{\nano\meter}, and can differ from fiber to fiber. During the period of measuring over the deployed link in 2023, several parts of the fiber were adapted and re-routed as part of restructuring of the network by KPN, changing both their length ($\pm$ \SI{100}{\meter}) and loss ($\pm$ \SI{1}{\decibel}) multiple times over the span of a few weeks. Table~\ref{table:fibers} shows typical loss values measured during this period. Thanks to the modular approach of our system, we have complete freedom in changing the arrival time of the photon in the midpoint, allowing for easy re-alignment of the propagation delay, both for the fiber restructuring as well as the daily expansion and contraction.

\begin{table}[]
\label{table:fibers}
\caption{\textbf{Parameters of Quantum Link.}}
\begin{tabular}{| c |M{2cm}| M{2cm} | M{1.5cm} | M{2.2cm} | M{2.2cm} |}
\hline
Link                        & Loss local & Loss deployed fiber & Total & Typ. Round-trip time, drift\\ \hline
The Hague to midpoint & \SI{3}{\decibel} QFC, \SI{4.9}{\decibel} filters    & \SI{5.2}{\decibel}, \SI{0.51}{\decibel\per\kilo\meter}   & \SI{13.1}{\decibel}           & \SI{103.16}{\micro\second} $\pm$\SI{40}{\pico\second\per\hour}\\ \hline
Delft to midpoint & \SI{4}{\decibel} QFC, \SI{4.5}{\decibel} filters    & \SI{5.6}{\decibel}, \SI{0.39}{\decibel\per\kilo\meter}   & \SI{13.9}{\decibel}             & \SI{145.31}{\micro\second} $\pm$\SI{80}{\pico\second\per\hour}\\ \hline
\end{tabular}
\end{table}

\section{Calibrations}
\subsection{Entanglement generation}
We employ a method where we interleave the generation of entanglement with three calibrations that measure and re-calibrate critical parameters and settings. These calibrations are an adapted version of the entanglement generation sequence that measure specific key elements of the entanglement generation: the signal-to-noise ratio (SNR), optical phase stability (PHASE) and the entangled state phase (XsweepX). Typical results of these three calibrations are shown in Fig.~S2. When these calibrations have passed their threshold, we proceed to measure the correlators used to calculate the entangled state fidelity, FID for short.

The first calibration is the measurement of the single-photon signal from both nodes and the background, which includes all the losses that are present in the system. We prepare the communication qubit in the $\ket{0}$ to maximize the single photon signal.
The measurement is the same experimental sequence as shown in Fig.~S4A, where now the state generation $\alpha$ pulse is removed to measure the signal, and replaced by a $\pi$-rotation to measure the noise. Furthermore, the arrival time of the photons is shifted by \SI{100}{\nano\second} to measure the brightness of both nodes individually.

Second is determining the contrast and phase of the relative optical phase between the nodes. We realize this by interfering a bright pulse (\SI{1}{\micro\second}) of the excitation lasers which we reflect off the diamond surface, and follows the exact path as the NV-center emission from there. By varying the setpoint of the phase in the stabilization scheme, we can measure a full interference fringe at the SNSPDs as seen in Fig.~S2B and characterize the residual phase noise of the system at the specific time window where single photons interfere.

Furthermore, this calibration is used to set the phase of the optical interference at any point along this fringe, which can be used to optimize the entangled state fidelity~\cite{Hermans_singleclick_2023}. We show the capability of changing the optical phase to two different setpoints of \SI{0}{\degree} and \SI{180}{\degree}.

The third and last calibration is the measurement of the entangled state phase, which has an offset with respect to the optical phase setpoint. Using the single-click protocol we create a correlated spin-spin state that we can use to probe this phase. We do this by measuring correlations in the rotated basis $X$ on Node Delft, while varying the readout basis in the Bloch sphere equator plane for Node The Hague. We can extract this entangled state phase by fitting the resulting oscillation with a single cosine, see Fig.~2C. We feed-forward this phase in the entanglement measurements to always generate the $\Psi^{\pm} = \frac{1}{\sqrt{2}}\left(\ket{01} \pm \ket{10}\right)$ during the entanglement generation.

\subsection{Experimental setup performance}
We maintain similar performance parameters of the experimental setup by performing calibrations of many different settings at different timescales. On a daily timescale, performance parameters of the communication qubit are re-calibrated: 
\begin{enumerate}
    \item Laser power (see Fig.~S10)
    \item Single-Shot Readout
    \item Microwave qubit control ($\pi$, $\pi/2$, $\alpha$)
    \item Optical excitation (optical $\pi$ pulse) for entanglement generation attempt
    \item QFC efficiency 
\end{enumerate}

Roughly every 20 minutes, a set of shorter timescale calibrations are run. Some are dependent on performance parameters that are available live during or right after a measurement set. See Fig.~S10 for the flowchart showing the calibrations and decisions made. As discussed in the previous section, we require several types of measurements to calibrate the phase of the entangled state. If in any of those measurements we are outside of the bound of acceptance, we restart the sequence. 
Additionally, the FID measurement sequence is cut into smaller blocks of 10k (30k) successful CR-checks for the heralded (delayed-choice) metropolitan distance experiments. After each block, the CR-check passing rate is checked against the average CR-check rate during the SNR sequence (per node). If the passing rate is below this value, we exit the FID sequence and continue. We do this because the CR-check passing rate is a value that includes many derived underlying performance parameters of the setup as a whole, e.g. laser wavelengths, laser power, objective position and others. If any of those values changed significantly the CR-check passing rate is negatively impacted, which we use as a signal that an underlying parameter of the experiment is drifting off ideal and subsequently we break out of FID data taking. As can be seen in the flowchart, afterwards all calibrations are restarted, and we start from a calibrated setup with the SNR sequence, which is when we expect the highest CR check passing rate, and as such, use that to define our threshold for the rest of the experiment.

\section{Modeling the generated entangled state}
To assist in setting up the experiment and simulating the outcomes, we employ a Monte-Carlo simulation based on the model described in~\cite{Hermans_singleclick_2023}. This is an extensive description of generating entanglement between remote nodes using the single-click protocol. It considers the spin-spin density matrix, based on the detection events where at most two photons reach the central detectors. By calculating the probability of each detection pattern and its corresponding spin-state, it arrives at the average density matrix given non photon-number resolving detectors. It takes into account many physical parameters that have an impact on the entangled state fidelity, such as photon loss, photon indistinguishably, and the photon wavepacket shape. It also includes several experimental parameters such as the effects of residual phase noise, dephasing noise and darkcount probability.\\

In the experiments presented in the main text we do not apply a post-processing correction using the Charge-Resonance check after the readout of the spin-state. This correction can be done to identify sequences where one of the nodes is in the NV$^{0}$ state, and no longer optically responsive to the excitation laser, but only in a post-processed fashion. To accurately model the effect of including events when the NV-center is ionized on the fidelity with the ideal state, we extend the model by including the probability of one of the nodes being ionized. We assume that, when an NV photon from a particular node in the midpoint is detected, this node can not be in the ionized state, but the other node can be with a probability $p_{ion}$. If the event in the midpoint is due to a noise photon, either of the nodes can be in the ionized state. This probability is measured by saving the CR-outcomes after the readout sequence. We include this ionized state as a third level of our qubit, that is outside the space addressed by our microwave control. When applying a readout pulse however, it will be read out as the $\ket{1}$ ($=$dark) state. The single shot readout (SSRO) fidelity of the solid-state NV qubits are given in Table~\ref{suptab:model}. State readout correction is performed according to iterative Bayesian unfolding as described in~\cite{pompili_thesis}.\\

An overview of all the simulation parameters used and their measured/used value for the simulation is given in Table~\ref{suptab:model}. We have also simulated near-term performance of the same link based on SnV centers (keeping the collection efficiency the same), which have 16 times higher coherent photon emission probability per optical excitation, full use of the improved QFC technology and improved phase stability, shown in Table~~\ref{suptab:modelimproved}.

\begin{table}
\caption{\textbf{Error budget of deployed link post-selected and heralded entangled states.} Parameters noted with * are averages of data measured between the fidelity measurements, other are calibrated in separate experiments. The probability to detect a photon is given for $\alpha = 1$. If only one value is mentioned it is considered equal for both nodes. The infidelity contribution is given as if it was the only error present, to make direct comparison easier. Note that due to the large contribution of noise counts, the infidelity of the final simulated state with all the contributions taken into account does not follow the sum of the individual contributions.\label{suptab:model}}
\vspace{1cm}
\centering
\begin{tblr}{
  cells = {c},
  cell{1}{2} = {c=2}{},
  cell{1}{4} = {c=2}{},
  cell{3}{3} = {r=3}{},
  cell{3}{5} = {r=3}{},
  cell{9}{3} = {r=2}{},
  cell{9}{5} = {r=2}{},
  vlines,
  hline{1-3,7-9,12-19} = {-}{},
  hline{4-6,10-11} = {1-2,4}{},
}
Simulation parameter                     & Delayed-choice   &                 & Fully heralded   &            \\
                                         & Value            & Error con.      & Value            & Error con. \\
Det. prob.* (Delft, DH)[]                & 10.6e-6, 8.4e-6  & 32.7\%          & 10.0e-6, 7.1e-6  & 28.5\%     \\
$\alpha$* []                             & 0.25 and 0.25    &                 & 0.25 and 0.25    &            \\
Backgr. counts* (Det1, Det2) [Hz]        & 40.3, 42.8       &                 & 23.8, 22.0       &            \\
Double excitation probability []         & 0.04, 0.1        & 3.7\%           & 0.04, 0.1        & 3.8\%      \\
Res. phase noise std* [$^\circ$]         & $35$             & 9.8\%           & $45.3$           & 14.3\%     \\
Dephasing probability []                 & 0.01, 0.01       & 2.3\%           & 0.02, 0.04       & 4.5\%      \\
Spectral diffusion FWHM [MHz]            & 13               & 4.5\%           & 13               & 4.8\%      \\
Beam splitter imperfections []           & 0.95             &                 & 0.95             &            \\
Detection window length [ns]             & 10               & -               & 15               & -          \\
Optical Excitation Rabi angle [$^\circ$] & 150, 150         & -               & 150, 150         & -          \\
Ionization probability* []               & 0.035, 0.045\%  & 5.4\%           & 0.046            & 6.2\%      \\
PSB collection efficiency []             & 0.1, 0.1         & -               & 0.0, 0.0         & -          \\
Simulated error (all params)             & -                & 43.1\%, 42.2\%  & -                & 45.6\%     \\
Simulated Fidelity                       & 0.568,  0.576    & -               & 0.544            & -          \\
Measured error                           & -                & 43.2\%,  42.4\% & -                & 46.6\%     \\
Avg. SSRO fidelity                       & 95.14\%, 94.49\% & -               & 94.13\%, 94.98\% &            \\
\end{tblr}
\end{table}

\begin{table}
\caption{\textbf{Error budget of near-term and future improvements.} For the near-term performance, we simulated an improved ZPL fraction of the SnV, the implementation of the improved QFC on both nodes, and an improvement of the local phase-stabilization. For future improvements on the system with we have added improvements in qubit coherence via dynamical decoupling, reduction of double excitation via shorter optical excitation,reduced ionization by better control of the charge state of the emitter and improved mode overlap at the central beamsplitter. \label{suptab:modelimproved}}
\vspace{1cm}
\centering
\begin{tblr}{
  cells = {c},
  cell{1}{2} = {c=2}{},
  cell{1}{4} = {c=2}{},
  cell{3}{3} = {r=3}{},
  cell{3}{5} = {r=3}{},
  cell{9}{3} = {r=2}{},
  cell{9}{5} = {r=2}{},
  vlines,
  hline{1-3,6-9,11-17} = {-}{},
  hline{4-5,10} = {1-2,4}{},
}
Simulation parameter                     & Near-term               &            & Future                  &            \\
                                         & Value                   & Error con. & Value                   & Error con. \\
Det. prob. (Delft, DH)[]                 & 16x10e-6, 16x10e-6      & 4.6\%      & 16x10e-6, 16x10e-6      & 4.6\%      \\
$\alpha$ []                              & 0.05 and 0.05           &            & 0.05 and 0.05         &            \\
Backgr. counts (Det1, Det2) [Hz]         & 5.0, 5.0                &            & 5.0, 5.0                &            \\
Double excitation probability []         & 0.04, 0.04              & 1.3\%      & 0.01, 0.01              & 0.3\%      \\
Res. phase noise std [$^\circ$]          & 15                      & 1.7\%      & 15                      & 1.7\%      \\
Dephasing probability []                 & 0.02, 0.04              & 3.0\%      & 0.01, 0.01              & 1.0\%      \\
Spectral diffusion FWHM [MHz]            & 13                      & 3.3\%      & 13                      & 1.3\%      \\
Beam splitter imperfections []           & 0.95                    &            & 0.99                    &            \\
Detection window length [ns]             & 15                      & -          & 15                      & -          \\
Optical Excitation Rabi angle [$^\circ$] & 150, 150                & -          & 150, 150                & -          \\
Ionization probability []                & 0.046                   & 4.7\%      & 0.01                    & 1.0\%      \\
PSB collection efficiency []             & 0.1, 0.1                & -          & 0.1, 0.1                & -          \\
Simulated error (all params)             & -                       & 17.2\%     & -                       & 10.0\%      \\
Simulated Fidelity                       & 0.828                   & -          & 0.90                   & -          
\end{tblr}
\end{table}

\clearpage

\begin{figure}
\includegraphics[width=\textwidth]{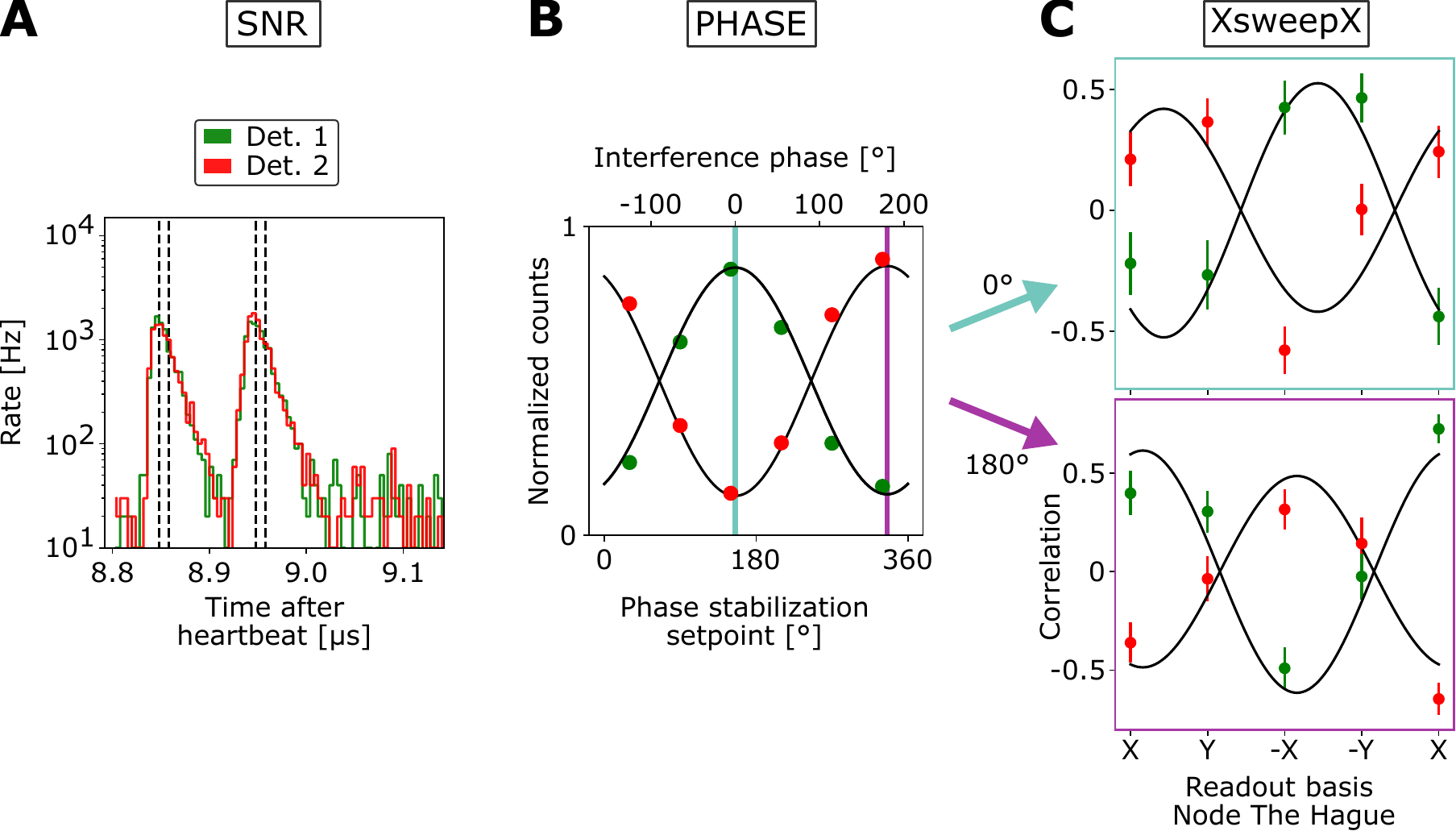}
\caption{\label{suppfig:fig2}Typical calibration outcomes of: \textbf{(A)} signal strength (SNR) where the dotted lines show the used `signal' window for both nodes respectively. The same windows are used to measure the 'noise' when the NV qubit state is rotated to the $\ket{1}$ (=dark) state.\textbf{(B)} Optical phase stability (PHASE), with fitted contrast (black lines). The vertical lines project to the top horizontal axis to determine the interference phase for the following measurements.\textbf{(C)} The entangled state phase (XsweepX) for an interference phase of \SI{0}{\degree} and \SI{180}{\degree}, showing the capabilities of the stabilization system to set a specific phase setpoint. The fitted phases (black solid lines) are 219(6) and 14(6) for the interference phase \SI{0}{\degree} and \SI{180}{\degree}, respectively.}

\end{figure}

\begin{figure}
\includegraphics[width=\textwidth]{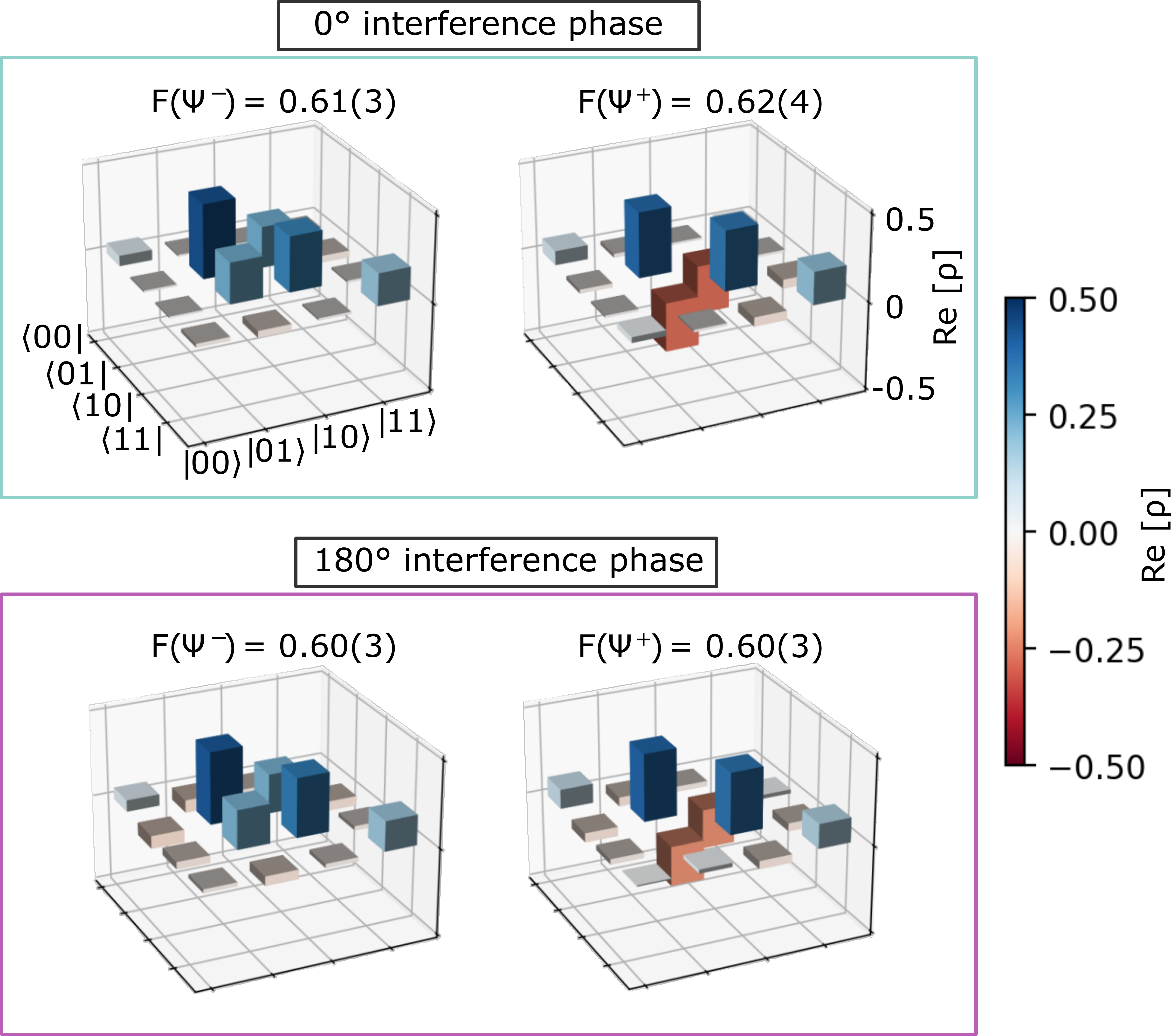}
\caption{\label{suppfig:fig3}Bayesian estimation of density matrix for both phase setpoints, using the post-selected entanglement generation scheme, with both nodes in the same lab in Delft. We measure all two-spin correlators $\langle M_{i}M_{j}\rangle$ with $ M_{i,j} \in \left[X, Y, Z\right]$ and use Bayesian estimation for tomography~\cite{Granade2016, Granade2017} to find the most likely density matrix. The calibration measurements in Fig.~\ref{suppfig:fig2} allow us to generate the same entangled state, at two different optical phase setpoints. The overlap with the ideal state for the most likely density matrix is given as numerical value, that is above $0.6$ for both phase setpoints and detector outcomes.}

\end{figure}

\begin{figure}
\includegraphics[width=1\textwidth]{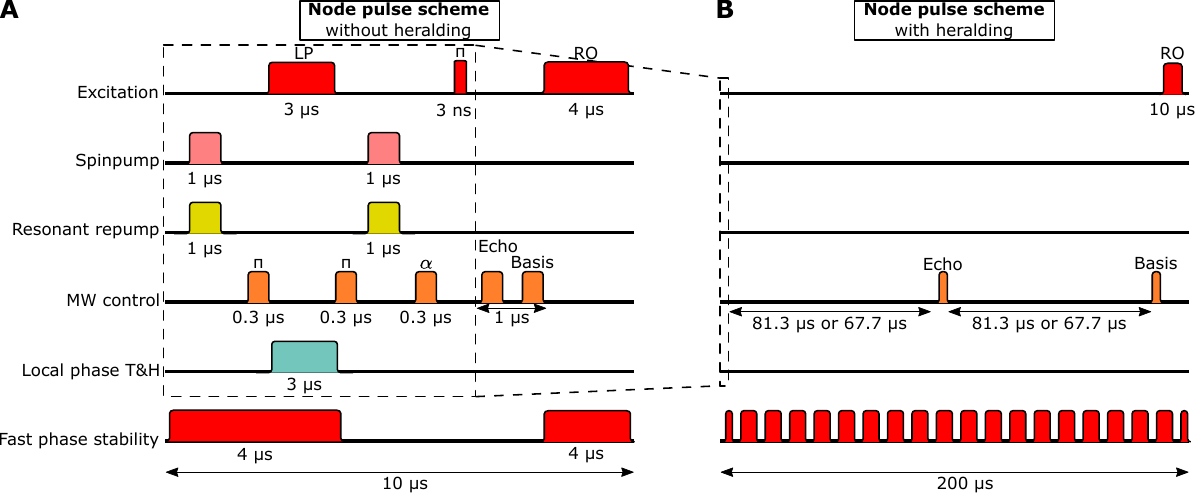}
\caption{\label{suppfig:fig4}\textbf{(A)} Local node pulse scheme for the duration of one heartbeat when performing entanglement generation without heralding. Red and yellow pulses are laser, orange are microwave pulses, blue-green is electrical feedback. The NV center is (re-)initialized in the correct spin and charge state via pumping resonant with $\ket{1}$ (\SI{637}{\nano\meter}) and the ZPL of NV$^0$(\SI{575}{\nano\meter}), respectively. We limit resonant optical excitation during local phase stability by preparing the qubit state in the opposite spin state a using microwave $\pi$-pulse. After the local phase stability, we initialize the spin into the $\ket{0}$ state, create the unequal superposition with a microwave pulse with rotation angle $\alpha$, and optically excite using a short optical excitation, completing the spin-photon entangled state generation. A microwave $\pi$-pulse echos the spin state, and a $\frac{\pi}{2}$-pulse selects the readout basis for the resonant optical readout that follows. \textbf{ (B)} Node pulse scheme during entanglement generation. This sequence now takes 20 heartbeat periods $\left(\SI{200}{\micro\second}\right)$. All pulses required to perform the local phase stability and spin-photon entanglement are repeated exactly as in (A). However, now the rephasing echo pulse is played at its calibrated value of \SI{82}{\micro\second}(\SI{68}{\micro\second}) for Node Delft (The Hague), while the stabilization light still is periodically sent to the midpoint such that phase stability feedback can be applied there. The total duration of \SI{200}{\micro\second} is for one whole entanglement generation attempt where wait time padding is added node specifically (not shown) to allow for a common repetition period when dealing with the different node-midpoint communication delays.}
\end{figure}

\begin{figure}
\centering
\includegraphics[width=0.75\textwidth]{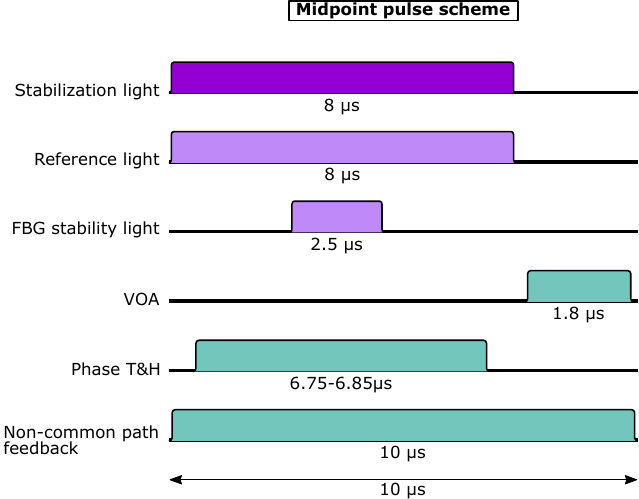}
\caption{\label{suppfig:fig5}Midpoint pulse scheme for the duration of one heartbeat. This sequence is agnostic to the experiment is run. Purple colors are telecom wavelength laser light, blue-green is electrical signal. Stabilization light is converted light coming from the nodes. The reference and FBG stability light are both originated from the telecom laser at the midpoint. The error signal for the fast lock is processed by a Track and Hold (T\&H) amplifier before being processed by the controllers. Variable optical attenuators are used to shield the detectors during the times where stabilization/reference light is propagating in the system, and they are only opened for \SI{2}{\micro\second} where single photons from the nodes can arrive. The global control task is not modulated, as the feedback bandwidth is much slower than the low-pass filtering of the error signal (\SI{150}{\hertz}). For all above pulse schemes holds that absolute block sizes are not to scale, the denoted time duration is leading.}
\end{figure}

\begin{figure}
\includegraphics[width=1\textwidth]{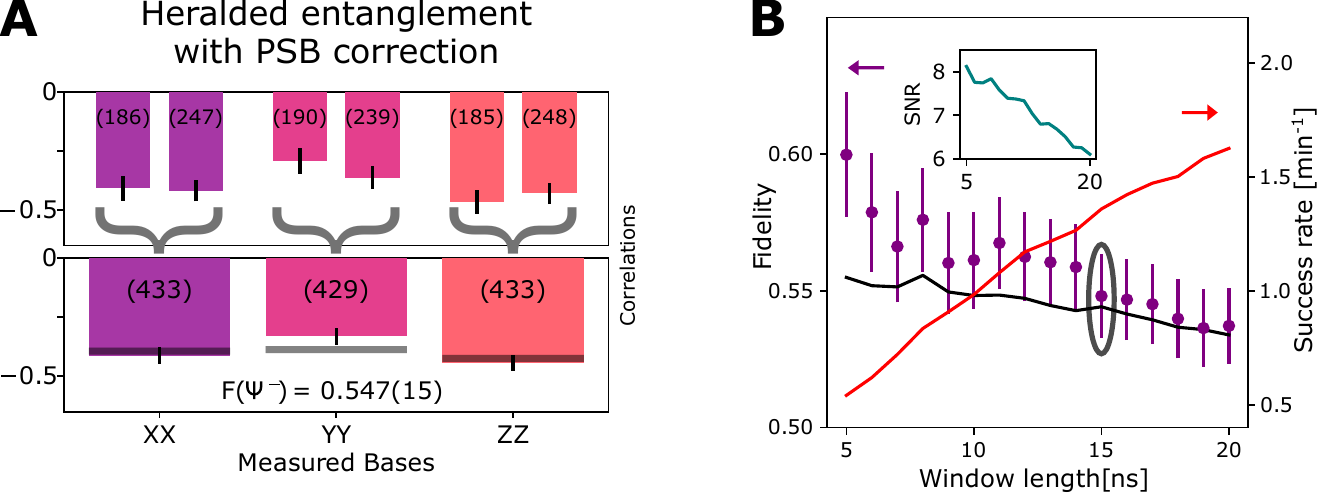}
\caption{\label{suppfig:fig6}\textbf{(A)} Correlation measurement for full heralding with correction on discarding events accompanied by a measured PSB photon, independent of detector outcome. Upper plot shows events per detector, lower for both detectors combined. Bars indicate measured data, the number in parenthesis indicate the amount of events. Horizontal lines indicate the theoretical model. \textbf{(B)} State fidelity and entanglement generation rate for varying photon acceptance window length. Inset shows signal-to-noise ratio for the same window lengths. All measurement outcomes are corrected for tomography errors, errorbars are 1 standard deviation.}
\end{figure}

\begin{figure}
\centering
\includegraphics[width=0.8\textwidth]{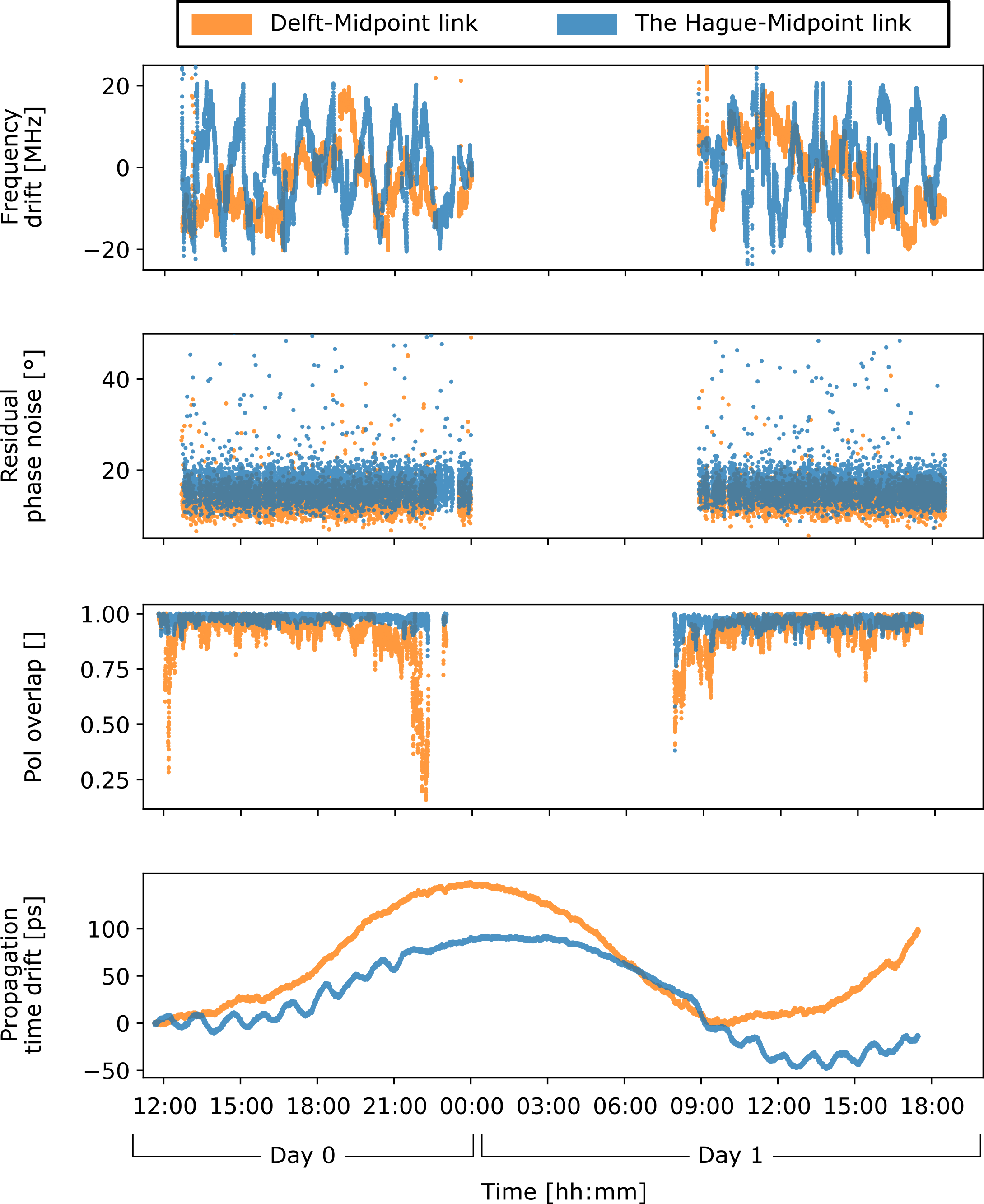}
\caption{\label{suppfig:fig7}Continuous data of important parameters over deployed fiber link. From top to bottom: the frequency drift, phase noise, polarization and time of flight as measured by the individual systems and logged into a central database. For the frequency, phase and polarization we have conditioned the data on the respective feedback system being active. The pause in the system overnight was due to one of the feedback systems being out of lock, which got reset the next morning.}
\end{figure}

\begin{figure}
\centering
\includegraphics[width=0.8\textwidth]{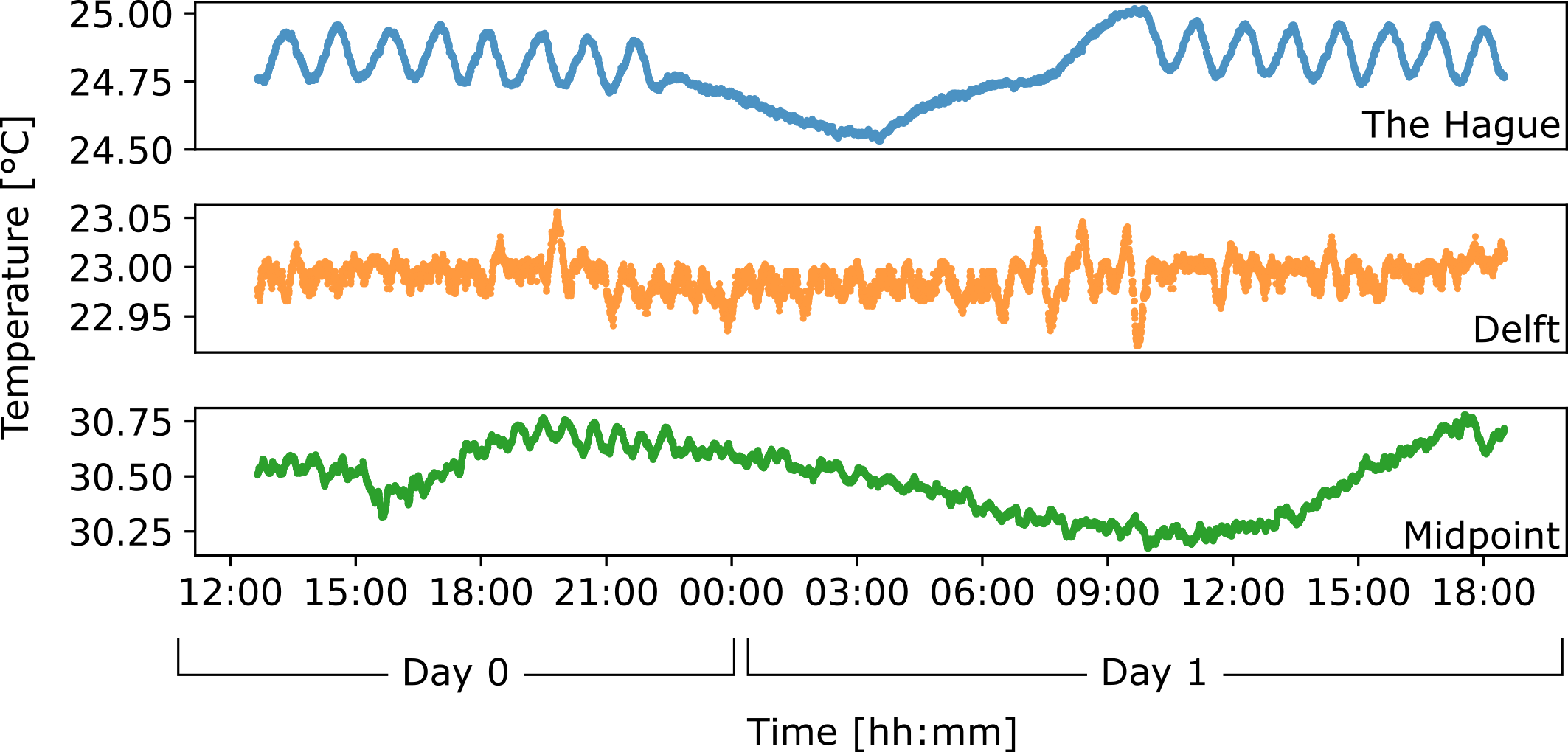}
\caption{\label{suppfig:fig8}Temperature logging of the three locations during the same time as the continuous data logging of Fig.~\ref{suppfig:fig7}.}
\end{figure}

\begin{figure}
\centering
\includegraphics[width=0.5\textwidth]{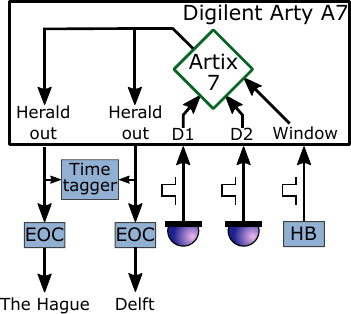}
\caption{\label{suppfig:fig9}Single photon detection signals are processed by custom FPGA-code on the Digilent Arty A7. If photons arrive within the externally derived acceptance window from the heartbeat (HB), a heralding signal is sent to both locations through an Electrical-to-Optical-Converter (EOC). The heralding signal is also recorded on the timetaggers on all the three locations.}
\end{figure}

\begin{figure}
\centering
\includegraphics[width=0.9\textwidth]{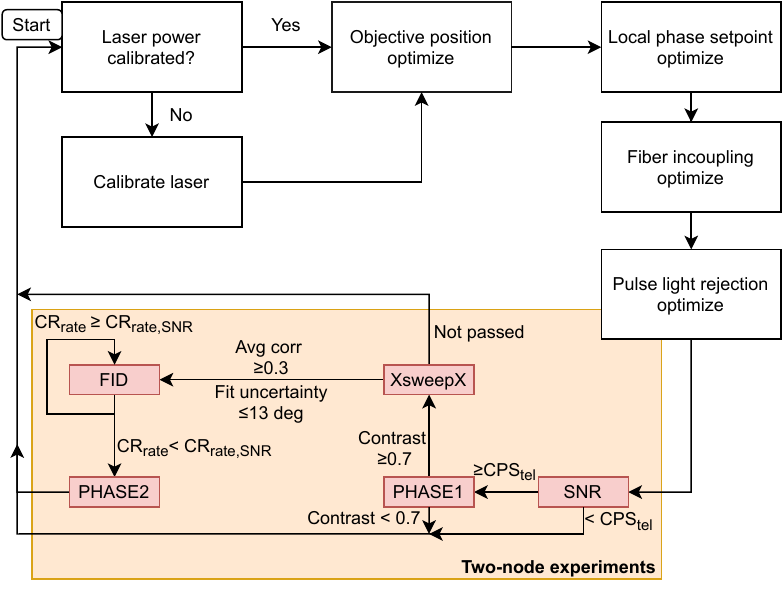}
\caption{\label{suppfig:fig10}Flowchart for calibrations and performing the sequence of two-node experiments for the purpose of measuring the entanglement fidelity of the entangled state that is aimed to be generated. Between the different steps of the measurement sequence thresholds are checked of measured parameters that decide continuation of the sequence. Calibrations are started with a signal-to-noise (SNR) measurement, where the counts-per-shot of telecom signal from the nodes $\left(\mathrm{CPS_{tel}}\right)$ is checked. We continue with PHASE and XsweepX and check their respective parameters. If all thresholds are satisfied, we continue with the fidelity measurements (FID) until the Charge Resonance passing rate $\left(\mathrm{CR_{rate}}\right)$ drops below the calibrated value it had during the SNR $\left(\mathrm{CR_{rate, SNR}}\right)$. Finally, we do another PHASE check before we restart the entire cycle.}
\end{figure}

\clearpage

\end{document}